\documentclass[aps,prb,reprint,superscriptaddress,showpacs,longbibliography,floatfix,footnoteinbib]{revtex4-2}
\usepackage{amsmath,amssymb,graphicx,units}
\usepackage[plainpages=false,pdfpagelabels,colorlinks=true,linkcolor=red,urlcolor=blue,citecolor=blue,pdftitle={Title},pdfauthor={},pdfdisplaydoctitle=true,pdfduplex=DuplexFlipLongEdge]{hyperref}
\usepackage{multirow}
\usepackage{setspace}
\usepackage{bm}
\usepackage{lipsum} 

\begin{document}

\title{Investigating Berezinskii-Kosterlitz-Thouless phase transitions in Kagome spin ice by quantifying Monte Carlo process: Distribution of Hamming distances}

\author{Wen-Yu Su}
\thanks{Those authors contributed equally to this work.}
\affiliation{Key Laboratory of Quantum Theory and Applications of MoE, Lanzhou Center for Theoretical Physics, and Key Laboratory of Theoretical Physics of Gansu Province, Lanzhou University, Lanzhou, Gansu 730000, China}

\author{Feng Hu}
\thanks{Those authors contributed equally to this work.}
\affiliation{School of Physics, Beihang University, Beijing 100191, China}

\author{Chen Cheng}
\email{chengchen@lzu.edu.cn}
\affiliation{Key Laboratory of Quantum Theory and Applications of MoE, Lanzhou Center for Theoretical Physics, and Key Laboratory of Theoretical Physics of Gansu Province, Lanzhou University, Lanzhou, Gansu 730000, China}

\author{Nvsen Ma}
\email{nvsenma@buaa.edu.cn}
\affiliation{School of Physics, Beihang University, Beijing 100191, China}

\begin{abstract}
We reinvestigate the phase transitions of the Ising model on the Kagome lattice with antiferromagnetic nearest-neighbor and ferromagnetic next-nearest-neighbor interactions, which has a six-state-clock spin ice ground state and two consecutive Berezinskii-Kosterlitz-Thouless (BKT) phase transitions. Employing the classical Monte Carlo (MC) simulations, the phases are characterized by the magnetic order parameter, and the critical temperatures are obtained by the finite-size scaling of related physical quantities. Moreover, we attempt to gain general information on the phase transitions from the MC process instead of MC results and successfully extract the correct transition points. Specifically, we focus on the selected data set of uncorrelated MC configurations and quantify the MC process using the distribution of two-configuration Hamming distances in this small data collection. This distribution is more than a quantity that features different behaviors in different phases but also nicely supports the same BKT scaling form as the order parameter, from which we successfully determine the two BKT transition points with surprisingly high accuracy. More strikingly, the distribution of Hamming distances can even determine the transition type without the help of the order parameter. We also discuss the connection between the phase transitions and the intrinsic dimension extracted from the Hamming distances, which is widely used in the growing field of machine learning and is reported to be able to detect critical points.
Besides providing a new understanding of the spin ice transitions in the Kagome lattice, we hope our proposal can be used to tackle the complicated phase transitions in the newly found compound HoAgGe and the quantum spin liquid state without a well-defined order parameter.
\end{abstract}
\maketitle

\section{Introduction}
\label{sec:Intro}

In a simple magnetic system, such as the two-dimensional Ising square lattice with antiferromagnetic nearest neighbor (NN) couplings, the ordered magnetic state at low temperatures changes into the paramagnetic state as temperature increases with a single finite-temperature phase transition characterized by the order parameters, corresponding susceptibilities or correlations. The universality class of the phase transition is predicted through the scaling behavior and critical exponents of those physical quantities~\cite{Sandvik2010}. However, the phases and transitions can be totally different once the geometric frustration is introduced, where the competing interactions that cannot be satisfied simultaneously bring degeneracy and result in exotic ground states in both quantum and classical cases~\cite{balents}. In classical phase transitions, as temperature increases, the complex ordered ground state usually turns into the disordered phase at high temperatures step by step through several intermediate phases rather than one direct phase transition. These consecutive phase transitions and different kinds of ordered or partly-ordered intermediate states are observed in the various frustrated spin systems and attract many theoretical and experimental investigations~\cite{KZhao,perron,chern,prlchern,Moessner,Haidong,wan}. On the other hand, these intermediate phases are ordinarily less known and sometimes even without a well-defined order parameter, make it difficult to understand the phase transitions in frustrated spin systems. What's more, in many theoretical and experimental studies, the existence of the intermediate phases is controversial in aspects of traditional physical properties similar to the ones used in simple Ising magnets, such as order parameters, susceptibility or correlations~\cite{guojing,liulu,Wangling,Ferrari2020,Nomura2021,Liu2022}.

In recent years, exploring new methods to detect phase transitions with less prior knowledge of the phases, especially integrating the machine learning (ML) ideas with the Monte Carlo simulations, has attracted increasing interest~\cite{melko,ising2,ising3,xy1,xy3,xy2,santos1,id2,xy4}. The ML techniques can be used to directly recognize the differences between phases and trace the critical points by classifying these phases, and have succeeded in detecting the second-order phase transitions~\cite{melko,ising2,ising3} and Berezinskii-Kosterlitz-Thouless (BKT) phase transitions~\cite{xy1,xy3,xy2} in some well-studied spin systems. While most of these works still require prior knowledge as the ML approach is based on the numerically obtained observables from MC simulations~\cite{melko,ising2,ising3,xy1,xy3}, some recent studies have attempted to get the universal information of phase transitions from analyzing only the raw data of spin configurations generated through MC processes~\cite{xy2,xy4,santos1,id2}. In principle, the latter approaches, which do not depend on the specific physical quantities and details of the target system, can hopefully inspire further works with the same idea to probe phase transitions universally by quantifying the MC process.

More than classifying the phases and locating the transition points, recent works demonstrate that the MC process can further verify the universality class of the phase transition and extract the corresponding critical exponent via finite scalings~\cite{santos1,id2}. In some sense, the MC sampling in the huge configuration space shares a similar idea with the ML procedure, which identifies the universal property of the high-dimensional data sets from minimally processed data collection. The key aspect is to extract the concerned information with finite degrees of freedom on the system with seemingly infinite complexity. Enlightened by the ML studies and along the same line, we aim to find a new way independent of models or universality classes that can describe the phase transition by quantifying the MC process rather than physical quantities and without employing any specific ML techniques. 

On the other hand, our work is in parallel inspired by recent studies on the dynamical phase transition in the context of thermalization and many-body localization (MBL) in closed quantum systems~\cite{Nandkishore2015,MBL_RMP2019}. The disorder-induced thermal-MBL transition can be characterized by analyzing the Hamming distances between the Fock states in the configuration basis, with the probability of Fock states determined by the target wavefunction~\cite{Hauke2015,Smith2016,Guo2021,Tomasi2021,Yao2023}. Extending the similar idea to the MC procedure, where the partition function determines the distribution of MC thermal configurations, we presume that the distribution of Hamming distances between these sampled configurations can possibly probe the phase transitions. In this work, we adopt this conception to reinvestigate the phases and phase transitions in the frustrated kagome Ising model (KIM) with antiferromagnetic NN couplings and ferromagnetic next nearest neighbor (NNN) ones, as shown in Fig.~\ref{fig:lattice}(a). The system has two consecutive temperature-driven BKT phase transitions with the charge-ordered spin-ice state of six-fold symmetry at low temperatures~\cite{takagi,chern,wills}. The same ordered ground state has recently been realized in the inter-metallic compound HoAgGe~\cite{KZhao}, supporting the naturally existing kagome spin ice for the first time. We expect the present work to reveal the rich physics in this system via quantifying the MC procedure without any physical quantities. 

The rest of the paper is organized as follows. In Sec.~\ref{sec:model}, we introduce the Kagome spin ice model and BKT phase transitions therein, which have been systematically reinvestigated by MC simulations according to physical quantities in Sec.~\ref{sec:obs}. In Sec.~\ref{subsec:PD}, we demonstrate that the critical points can be retrieved by scaling with BKT form using non-physical quantities, specifically the distribution of Hamming distances of the selected configurations in the MC procedure. The intrinsic dimension, which is commonly used in ML in detecting transitions, is also discussed in Sec.~\ref{subsec:Id}. In Sec.~\ref{sec:generity_PD}, we propose that the type of the phase transition can be determined by the distribution of Hamming distances even without the help of the order parameter.  Finally, the summary and discussion are presented in Sec.~\ref{sec:sum}.

\section{Preliminaries}
\label{sec:model}

\begin{figure}[!t]
\includegraphics[width=\columnwidth]{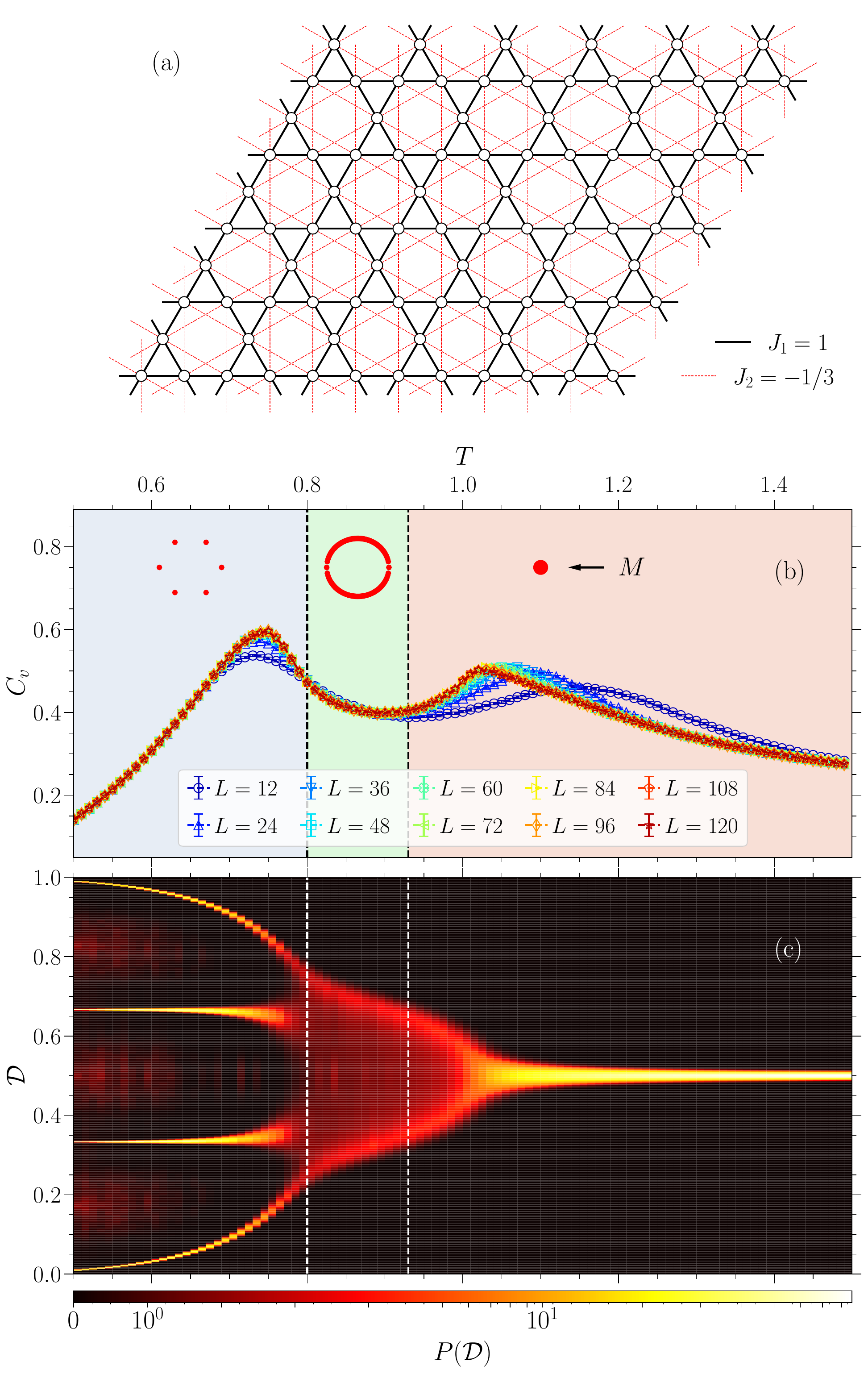}
\caption{(a) Ising model on the Kagome Lattice with antiferromagnetic/ferromagnetic interactions between the NN/NNN sites. The lattice constant $a=1$ measures the distance between the nearest unit cells. (b) Specific heat $C_v$ as a function of temperature $T$ for different $L$. Here and after, the error bar of MC obtained data depicts the statistical error in Eq.~\eqref{eq:ave}. The schematic plot of order parameter $M$ in the complex \{Re$M$, Im$M$\} plane demonstrates the three phases: the six-state clock spin ice, the intermediate critical region, and the disordered phase from left to right [See Appendix~\ref{sec:M_hist} for the numerically obtained $M$]. (c) The distribution of Hamming distances $P({\cal D})$ for $L=120$ displayed in the 2-D parameter space $\{ T, {\cal D} \}$. The vertical dashed lines in (b) and (c) depict two critical points, the values of which are determined by the scaling of $M$ [See Fig.~\ref{fig:scaling_MXM}].}
\label{fig:lattice}
\end{figure}

We study the two-dimensional Ising model on the kagome lattice with antiferromagnetic NN interactions and ferromagnetic NNN ones. The model Hamiltonian reads  
\begin{align}
H=J_1 \sum_{\langle ij\rangle} \sigma_i \sigma_j + J_2 \sum_{\langle\langle ij\rangle\rangle} \sigma_i \sigma_j
\label{eq:ham}
\end{align}
where $\sigma_{i}=\pm 1$ denotes the Ising spin on the site $i$, and $J_{1}>0$ ($J_{2}<0$) is the interaction between the NN (NNN) sites on the Kagome lattice illustrated in Fig.~\ref{fig:lattice}(a). The KIM described by Eq.~\eqref{eq:ham} features the ground state of a six-state clock spin ice~\cite{chern,takagi}, which can be described by a complex order parameter
\begin{align}
M=\frac{1}{N}\sum\sigma_{i}{\rm exp}(\mathrm{i}\bm{Q}\cdot\bm{r}_{i})
\label{eq:op}
\end{align}
with $\bm{Q}=(4\pi/3,0)$ and $N$ for the total number of spins. In the ordered phase at low temperatures, the order parameter follows $M=|M|{\rm e}^{\mathrm{i}\phi}$ with magnitude $|M|\neq 0$, and the phase $\phi$ can only take one of the values from $\phi=n\pi/3$ with $n=0,1,2,3,4$ and $5$, as illustrated in Fig.\ref{fig:lattice}(b). As temperature increases, the system first goes into a critical phase with power-law decay of spin correlations and unrestricted values of $\phi$, which exhibits an emergent U(1) symmetry~\cite{chern}. At sufficiently high temperatures, all orders break down, and the system features a disordered state. The KIM undergoes two finite-temperature phase transitions, which belong to the same universality class as the six-state clock model~\cite{Wolf}. Both phase transitions in the $q$-state clock model are predicted to belong to the Berezinskii-Kosterlitz-Thouless (BKT) universality class as long as $q\geqq 5$ from theoretical analysis~\cite{Landau}. Although some numerical studies claim that BKT transitions only occur for systems with $q\geqq 8$~\cite{carlos,carlos2}, most recent works agree with the idea that those two phase transitions in $q=6$ clock model are of the BKT type through various computational methods~\cite{xiangtao,youjin,Surungan}. 

The six-clock spin-ice ground state has recently been observed in the inter-metallic compound HoAgGe for the first time in Ref.~\cite{KZhao}, where the authors have claimed that the physics in the experiment can be described by a generalized KIM with geometry distortions and weak dipolar interactions. Our present work aims to reinvestigate the physics of the known physics in the standard KIM in a universal way, and future works following the same procedure would help to understand phase transitions in more complex models without obvious order parameters, such as the compound HoAgGe. 

\subsection{MC Simulation of Physical Quantities}
\label{subsec:pre_pq}

This paper presents a large-scale computational study on the KIM in Eq.\eqref{eq:ham} with fixed $J_{2}=-1/3$ and $J_1=1$, with the latter as the energy scale. Specifically, we consider a system with $L\times L$ unit cells that contribute a total number of spins $N=3L^{2}$, and use the periodic boundary conditions to minimize the boundary effect. We adopt the classical MC method with a standard Metropolis sampling algorithm of single-spin update, where a randomly chosen spin is flipped with a probability $p={\rm exp}(-\beta\Delta E)$. Here $\Delta E$ is the energy change of the flipping, $\beta=1/k_{B}T$ with $T$ the simulated temperature, and $k_{B}$ is set as $1$ for simplicity. 

For all calculations with different system sizes ($L=12$ to $120$), we take $N_b=56$ independent bins of MC procedures~\footnote{Here $N_b=56$ equals the number of cores in the computing cluster for simulations in this work. Our results have confirmed that 56 bins are rational and sufficient. }; Each bin has 10000 MC steps to reach equilibrium and 20000 MC steps for measurement. These settings are sufficient to avoid the autocorrelation between the two MC steps. We thus employ a total number of $\propto 10^{6}$ configurations to get the expected thermodynamic averages for different observables. For a generic quantity $Q$, the value and statistical error are computed as 
\begin{align}
\bar{Q}&=\frac{1}{N_{b}}\sum_{b=1}^{N_{b}}\bar{Q_{b}}\\
\sigma_{Q}&=\sqrt{\frac{1}{N_{b}(N_{b}-1)}\sum_{b=1}^{N_b}(\bar{Q_{b}}-\bar{Q})^{2}}
\label{eq:ave}
\end{align}
with $\bar{Q}_{b}$ the average over all MC steps in each bin. Therefore, the estimated value for the expectation of $Q$ becomes $\langle Q\rangle=\bar{Q}\pm \sigma_{Q}$. One of the most often studied phase quantities in probing thermal phase transition is the specific heat $C_{v}$ defined as 
\begin{align}
C_{v}=\frac{1}{Nk_{B}T^{2}}(\langle E^{2}\rangle-\langle E\rangle^{2}),
\label{eq:spch}
\end{align}
where $E$ is the total energy of the system. For the conventional phase transitions, the peak value of $C_{v}$ as a function of temperature $T$ changes dramatically with the system size, indicating the singularity behavior at $L\to \infty$, which can be used to detect phase transitions. However, this is not the case for the BKT transition in KIM, where the peak height does not change much as $L$ increases, as shown in Fig.~\ref{fig:lattice} (b). Actually, the peak positions for the $T$-dependence of $C_{v}$ are usually away from the real critical points in the BKT transitions~\cite{kt1,carlos,carlos2,Jan}, and it is not trivial to define a proper physical quantity with known critical behavior. As for KIM, we can use the order parameter defined in Eq.~\eqref{eq:op}; However, the quantity used to detect a BKT transition usually depends on the specific phase, which can be tricky to find in a system with computational difficulties and without preliminary knowledge.

\subsection{Collection of Uncorrelated MC Configurations and Hamming Distance}
\label{subsec:pre_hd}

A standard approach analyzes the phases and phase transitions based on the physical quantities obtained from numerical simulations. In this paper, we consider another possibility: Can one learn the same physics without borrowing the concept of any physical quantities? For the quantum MC studies with the severe sign problem, some recent works have tried to extract physics from the sign value itself and reported several successful cases where the transition can be tackled by the anomaly of either the sign or its derivatives~\cite{Mondaini2022_science,Tarat2022,Mou2022,Mondaini2022,yan2023universal}. This procedure is apparently unsuitable for the classical MC simulations without the obstacle of the sign. In the present work, we aim to extract information on transitions that happens in the infinite configuration space by quantifying the limited set of sampled configurations that can be used in different systems and transitions. 

The MC simulation performs an importance sampling in the configuration space $\{\vec{\sigma}\}$, and the visited probability of a configuration $\vec{\sigma}$ at certain parameters is determined by the partition function. Therefore, it is rational to assume that the collection of visited configurations in the MC sampling procedure is closely related to the system's physical properties. In the data set of uncorrelated configurations, we are interested in the normalized Hamming distance between two configurations $\sigma$ and $\sigma^\prime$ defined as 
\begin{align}
    {\cal D}(\vec\sigma,\vec{\sigma^\prime}) = \frac{1}{N}\sum_i (1/2-\sigma_i\sigma^\prime_i/2).
\end{align}
Obviously, ${\cal D}=0$ for the same configurations and ${\cal D}=1$ between two exactly opposite ones. For totally uncorrelated data collections, the Hamming distances follow a Gaussian distribution centered at ${\cal D}=1/2$. The Hamming distance is widely used in the dynamical phase transition from thermalization to many-body localization (MBL) in closed quantum systems~\cite{Hauke2015,Smith2016,Guo2021,Tomasi2021,Yao2023}. In detecting the thermal-MBL transition, the time evolution of $\cal D$ from a single initial state can quantify the ergodicity of the system. For the thermal phase transitions that focus on the thermal equilibrium state, we can compute Hamming distances between every two configurations in a small set of data collection. Unlike some very recent works studying the static phase transitions using the average of the Hamming distance~\cite{Mou2022,Yi2022}, we focus on its distribution of $\cal D$ since the average $\cal D$ is always 1/2 if the MC procedure is not trapped by the local minimums in KIM.

\begin{figure}[!b]
\includegraphics[width=\columnwidth]{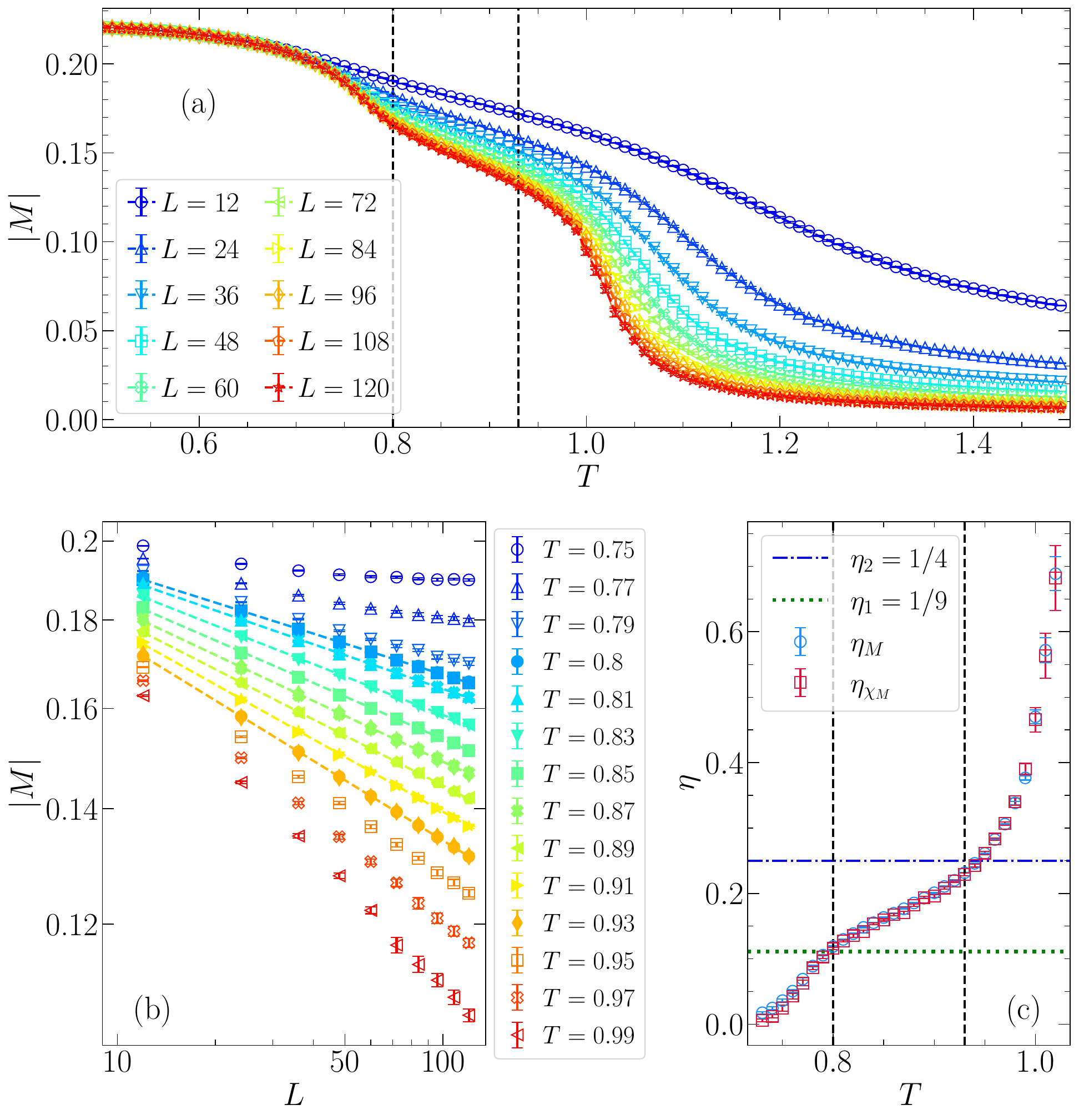}
\caption{ (a) The order parameter $|M|$ versus $T$ for different $L$. (b) $|M|$ versus $L$ at different temperatures, displayed in the log-log scale. The straight dashed lines depict power-law fitting in the intermediate critical phase. (c) The critical exponent $\eta$ extracted from the fitting $M\propto L^{-\eta/2}$ for different temperatures, where the error bar depicts the fitting uncertainty of $\eta$. The vertical dashed lines in (a) and (c) mark the critical points obtained from $M$ scaling. }
\label{fig:ML}
\end{figure}

To get a collection of uncorrelated configurations, we adopt $200$ equally-spaced configures in each bin and collect the data of all bins. With a total number $N_c\approx 10^5$, these configurations produce a data set of $\cal D$ with size $N_{\cal D}\approx 10^{10}$. This is a large number but still infinitely small compared to the total number of configurations $2^N$. Fig.~\ref{fig:lattice}(c) shows the distribution of the Hamming distance as a function of $T$ for $L=120$ as an example, where one can tell three different regions at the very first glance. At zero temperature, the Hamming distance between the six-fold degenerate ground states supports only four possible values (0, 1/3, 2/3, and 1) [See Appendix~\ref{appendix:six-clock}], which is consistent with the numerically obtained $P({\cal D})$ at $T=0.5$, where one observes a slightly broadening bright region instead of a delta function for the finite system at finite temperatures. As the temperature increases, the position of the $P({\cal D})=0$ (1) peak moves towards 0.5, as depicted by the bright curves, and the $P({\cal D})=1/3$ (2/3) peak hardly varies. These four sharp peaks disappear in the intermediate phase, where two broad distributions occur. Further increasing $T$ to sufficiently high temperatures, $P({\cal D})$ in the disordered phase has a symmetric Gaussian distribution centered at 0.5 as all configurations have equal probability. 

\section{BKT transitions in the aspect of physical quantities}
\label{sec:obs}

The BKT transitions in the $q$-state clock model as well as the KIM can be characterized by the order parameter $M$ and its susceptibility $\chi_{M}\equiv\frac{\mathrm{d}\langle M\rangle}{\mathrm {d}h}|_{h\to 0}$ to the magnetic field $h$, which is calculated using
\begin{align}
\chi_{M}=\frac{\beta}{N}(\langle M^{2}\rangle-\langle M\rangle^{2}). 
\label{eq:susc}
\end{align}
Then the critical temperature and exponents in the thermodynamic limit can be extracted from the standard finite-size scaling approach~\cite{Ma1,Aramthottil2021,Jan2020}. The order parameter and its susceptibility in KIM obey the scaling forms within the region close to $T_{c}$ as:
\begin{equation}
\begin{split}
    M=&L^{-\beta_c/\nu}{\cal F}_{M}(\xi/L)\\
\chi_{M}=&L^{\gamma/\nu}{\cal F}_{\chi}(\xi/L),
\end{split}
\label{eq:scaling1}
\end{equation}
where $\beta_c$, $\gamma$, and $\nu$ are the critical exponents for magnetization, susceptibility, and correlation lengths, with ${\cal F}_{M}$ and ${\cal F}_{\chi}$ are functional forms for the data of different system sizes $L$. For BKT universality class, the correlation length $\xi$ exponentially diverges at $T_{c}$ as $\xi\sim \rm{exp}(c/\sqrt{t})$ with $t=|T-T_{c}|$ and a non-universal constant $c$~\cite{kt}. Using the general exponent relations $\gamma=\nu(2-\eta)$ and $\beta_c=\nu(d-2+\eta)/2$ with the lattice dimension $d=2$, we can rewrite the scaling forms of $M$ and $\chi_{M}$ with a single scaling exponent $\eta$ as
\begin{equation}
\begin{split}
  M  = & L^{-\eta/2} {\cal F}_{M}^{-1} (L/e^{c/\sqrt{t}})\\
\chi_{M} =& L^{2 - \eta} {\cal F}_{\chi}^{-1} (L/e^{c/\sqrt{t}}).
\end{split} 
\label{eq:scaling2}
\end{equation}
Close to the critical points and inside the whole critical phase, the functional form ${\cal F}_{M}^{-1} (L/e^{c/\sqrt{t}})$ approximates to a constant value as $L/\xi\rightarrow 0$. Therefore $M$ (as well as $\chi_M$) should behave in a power law as a function of $L$ in the whole region between $T_{c1}$ and $T_{c2}$, as confirmed by the numerical results displayed in Fig.~\ref{fig:ML}(b), depicted by the linear behavior in the double logarithmic scale. Moreover, one can extract the critical exponent $\eta$ by fitting $M\propto L^{-\eta/2}$. 
The obtained $\eta$ is shown in Fig.~\ref{fig:ML}(c), where the exponents for both transition points and physical quantities ($M$ and $\chi_M$) agree well with the theoretical predictions, that is, $\eta(T_{c1})=1/9$ and $\eta(T_{c2})=1/4$~\cite{chern}. For the critical phase between $T_{c1}$ and $T_{c2}$, the order parameter is proportional to $\log L$ with slope $-\eta/2$ satisfying $1/9<\eta<1/4$.

\begin{figure}[!t]
\includegraphics[width=\columnwidth]{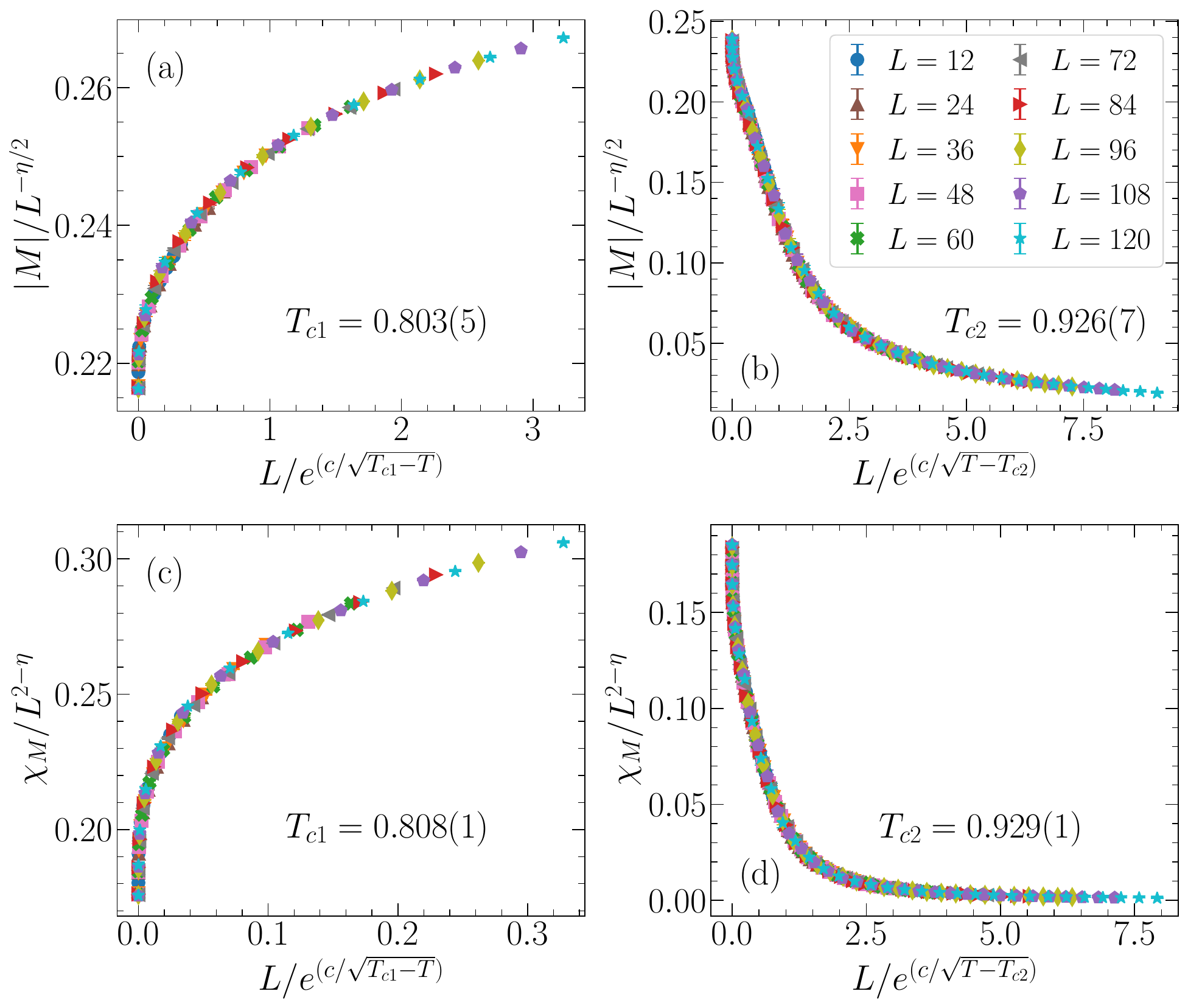}
\caption{The data collapse following the BKT scaling form~\eqref{eq:scaling2} for the two critical points and two physical quantities: $T_{c1}$ ($T_{c2}$) in left (right) panels and $M$ ($\chi_M$) in left/right panels. Here $\eta=1/9$ ($1/4$) is fixed in the scaling of $T_{c1}$ ($T_{c2}$), and the obtained critical temperature is shown by text in each panel. The non-universal constant $c$ = $1.16(6)$, $1.47(3)$, $1.94(8)$, $1.54(1)$ in panels (a-d), respectively. All panels share the same legend. }
\label{fig:scaling_MXM}
\end{figure}

We further carry out a standard finite-size scaling by the data collapse approach of the functional form in Eq.~\eqref{eq:scaling2}, as displayed in Fig.~\ref{fig:scaling_MXM}. In the scaling procedure, we adopted the above-mentioned $\eta$ as fixed values, and searching for the best data collapse is equivalent to the minimization problem in the two-dimensional parameter space $\{T_{c},c\}$ [See appendix~\ref{appendix:data_collapse} for the scaling detail and error estimation]. The results of $M$ scaling are shown in Fig.~\ref{fig:scaling_MXM}(a) and (b), where the data collection for different $L$ nicely collapse to a smooth curve, and the obtained critical points are in accordance with the fitting in Fig.~\ref{fig:ML}(b). Considering the error, the scaling of $\chi_M$ provides consistent results, as displayed by the corresponding data collapse in Fig.~\ref{fig:scaling_MXM}(c) and (d) for both critical points. Here and after, for all scaling procedures without specific instructions, the data collapse is carried out using the data within the temperature range [$0.70,T_{c1}$) for the first transition point, and ($T_{c2},1.25$] for the second.

It is worth noting that the critical points in the present work have non-negligible differences from what is obtained in Ref.~\cite{chern}, although the two works focus on the same model with the same parameters and numerical approach. The correctness of the present work and the origin of this discrepancy are discussed in appendix~\ref{appendix:benchmark}.

\section{Probing BKT transition by quantifying MC procedure}
\label{sec:HD}

We reinvestigate the BTK transitions in KIM with the finite-size scaling of the order parameter and its susceptibility in the previous section. %However, as we demonstrated in Sec.~\ref{sec:model}, not all physical quantities, such as specific heat, can correctly catch the transition points~\cite{xiangtao,youjin}. Moreover, the order parameter may be hard to define for the system without prior knowledge and has computational difficulties. 
In this section, we aim to study the phase transitions employing the universal quantities quantifying the MC procedure, which do not depend on the details of the target system. The main idea is to quantitatively analyze the MC visited configurations based on the Hamming distances in the small data set of selected uncorrelated configurations. 

\subsection{Distribution of the Hamming distance}
\label{subsec:PD}

\begin{figure}[!b]
\includegraphics[width=1\linewidth]{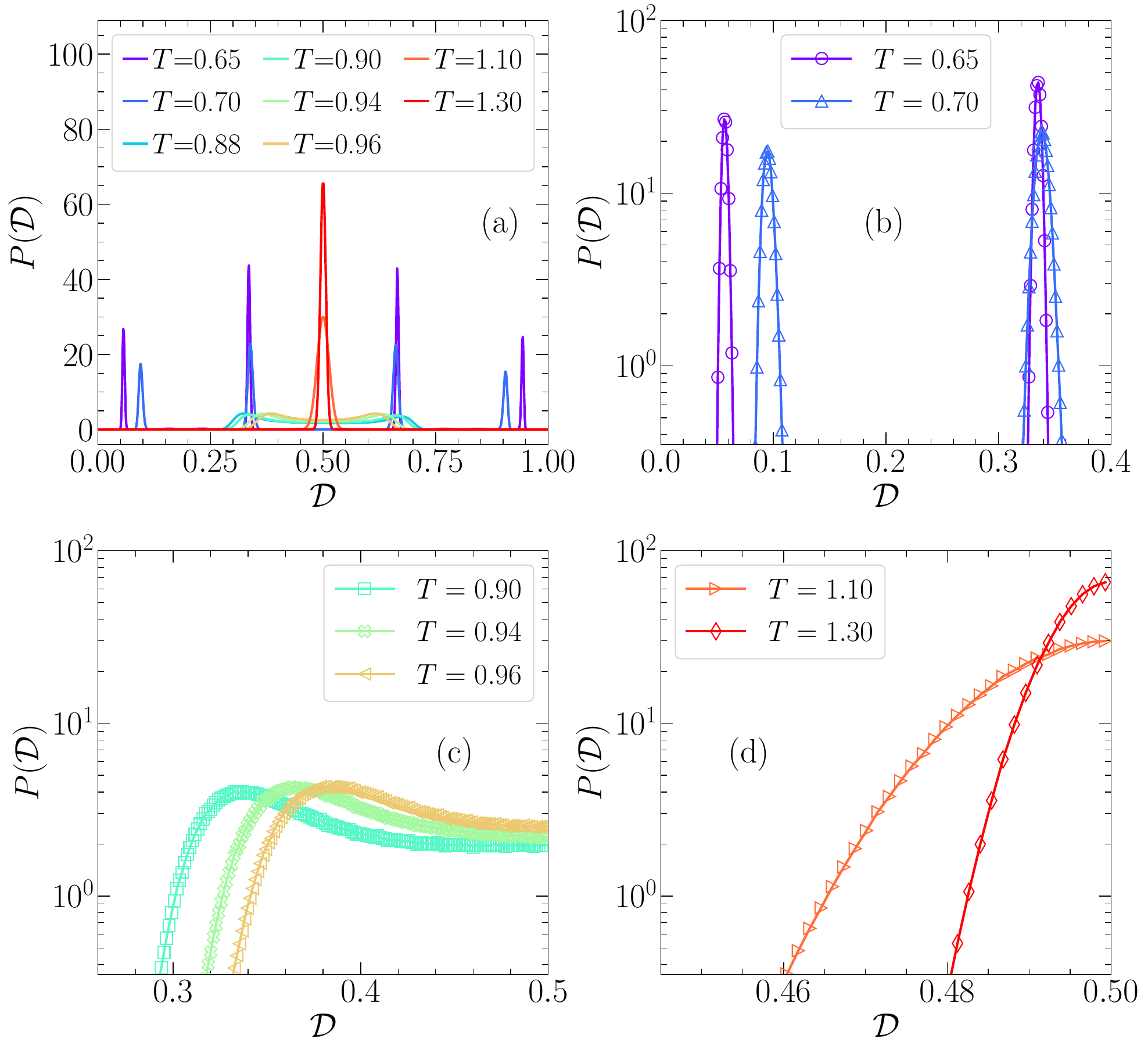}
\caption{(a) Distribution of Hamming distances $P({\cal D})$ for several temperatures in typical regions. Panels (b-d) show selected data in different phases and the Gaussian-like fitting of $P({\cal D})$; Markers and solid lines depict the original data and the fitting using Eq.~\eqref{eq:Gaussian_fitting}. For low temperatures in panel (b), the summation of two Gaussian forms catches two peaks; For the intermediate critical phase in panel (c), the two Gaussian forms adjust the shape of the peak; For higher temperatures ($T>1$) in panel (d), only one Gaussian form is used. }
\label{fig:PD}
\end{figure}

The distribution of Hamming distance features distinguishing behaviors in three phases, as demonstrated in Fig.~\ref{fig:lattice} (c) and related text in Sec.~\ref{sec:Intro}. In the following, we do not rest content with qualitatively resolving the phases and further seek a proper scaling process with the $P({\cal D})$. As specifically displayed in Fig.~\ref{fig:PD}(a), the distribution of Hamming distance is symmetric around $0.5$, and the curves of $P({\cal D})$ are quite smooth. Therefore, we try to fit $P({\cal D})$ for ${\cal D}\in[0,0.5]$ at each fixed $T$ by the summation of two Gaussian forms as:
\begin{equation}
\begin{split}
{\cal G}({\cal D})=&\frac{A}{\Delta \sqrt{2\pi}}\mathrm{exp}[-(\frac{{\cal D}-{\cal D}_0}{\Delta})^2/2]\\
            &+\frac{A^\prime}{\Delta^\prime \sqrt{2\pi}}\mathrm{exp}[-(\frac{{\cal D}-{\cal D}_0^\prime}{\Delta^\prime})^2/2],
\label{eq:Gaussian_fitting}
\end{split}
\end{equation}
where $D_0$ and $D_0^\prime$ depict two peak centers with the restriction $D_0<D^\prime_0$, $\Delta$ and $\Delta^\prime$ are the relative widths, parameters $A_1$ and $A_2$ count weights of the two peaks with $A_1+A_2=1$. Noted that there can only be one peak in the region ${\cal D}\in[0,0.5]$ at large temperatures, we manually set $A^\prime=0$ for $T>1$. As shown in Fig.~\ref{fig:PD}(b-d), this fitting nicely catches the original data. 

\begin{figure}[!t]
\includegraphics[width=\columnwidth]{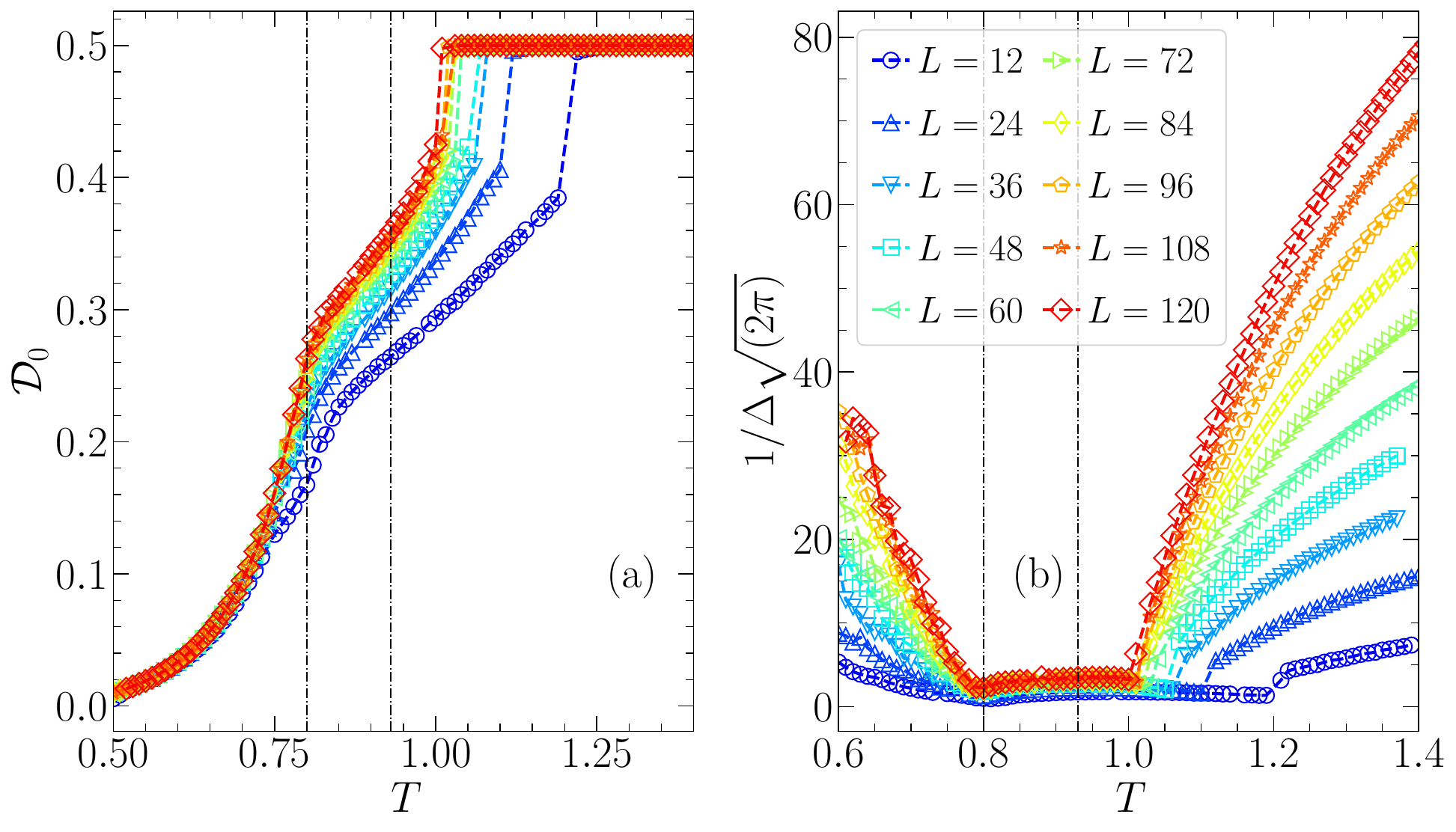}
\caption{The (a) position ${\cal D}_0$ and (b) effective height $1/\Delta\sqrt{2\pi}$ of the first $P({\cal D})$ peak versus temperature for different system sizes. }
\label{fig:Mu_A}
\end{figure}

\begin{figure}[!b]
\includegraphics[width=\columnwidth]{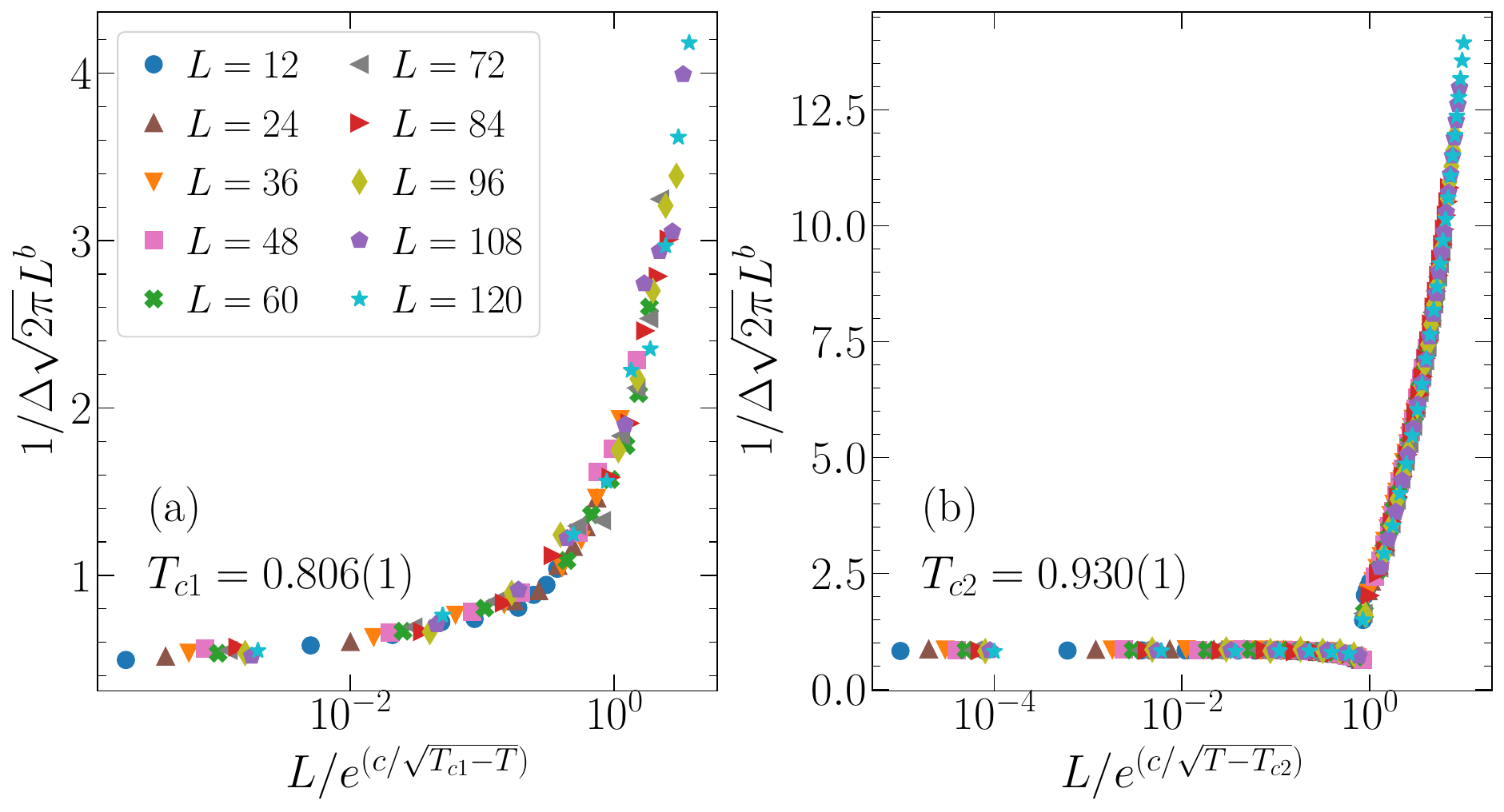}
\caption{Scaling of $1/\Delta\sqrt{2\pi}$ following the BKT scaling form~\eqref{eq:scaling_A} for (a) $T_{c1}$ and (b) $T_{c2}$, respectively. The obtained non-universal constants are: (a) $b=-0.27(1)$, $c=1.06(1)$; (b) $b=-0.29(1)$, $c=1.47(1)$.}
\label{fig:A_scaling}
\end{figure}

In the following, we focus on the evolution of the first peak, which starts from ${\cal D}_0=0$ at the zero temperature. The position $D_0$ and the effective height $1/\Delta\sqrt{2\pi}$ as a function of $T$ for various $L$ is displayed in Fig.~\ref{fig:Mu_A} (a) and (b), respectively. For both curves, there is an anomalous close to $T_{c1}$ and a discontinuous that is greater than $T_{c2}$. The latter rapidly decreases as $L$ increases, and its position is always the same in the two panels in Fig.~\ref{fig:Mu_A}. This information on possible transitions is similar to the order parameter $M$ and looks more prospective in detecting transitions than the specific heat. Therefore, we try a similar scaling procedure for the $1/\Delta\sqrt{2\pi}$ as the order parameter with a general BKT scaling form:
\begin{align}
    \frac{1}{\Delta\sqrt{2\pi}}L^{b}={\cal F}(L/e^{c/\sqrt{t}}),
    \label{eq:scaling_A}
\end{align}
where $b$ is the scaling exponent. Here we choose only $1/\Delta\sqrt{2\pi}$ for scaling since ${\cal D}_0$ is bounded up to 0.5. Different from the scaling of $M$ and $\chi_M$ where the exponent $\eta$ is known, here we have three parameters to be determined, and the searching for the best data collapse is carried out in the three-dimensional parameter space $\{T_c,b,c\}$. Nevertheless, the minimization process generates very good data collapse and critical points. As shown in Fig.~\ref{fig:A_scaling}, both critical points nicely agree with that obtained from the scaling of the order parameter and its susceptibility. 

\begin{figure}[!t]
\includegraphics[width=\columnwidth]{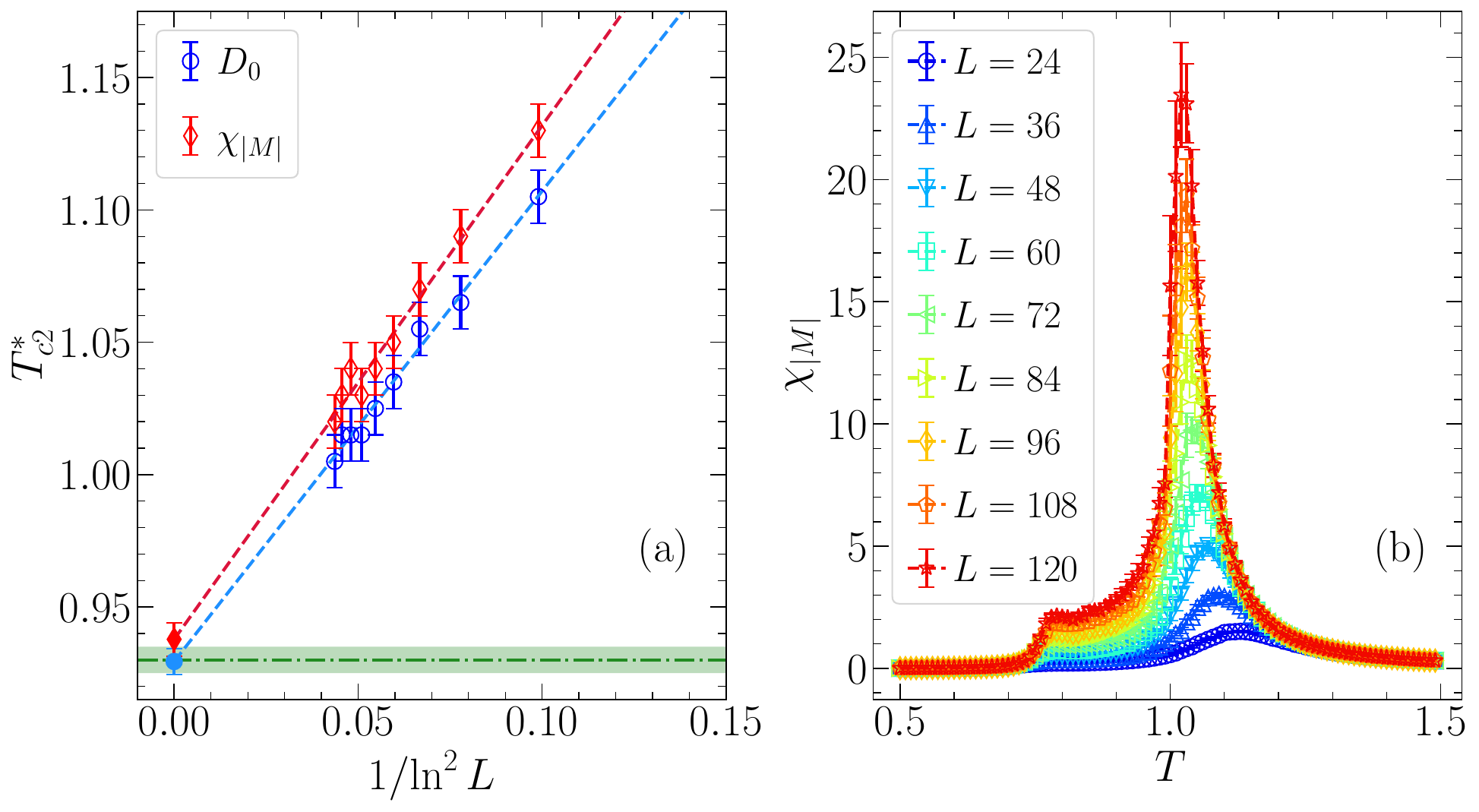}
\caption{(a) Finite-size extrapolation of $T^*_{c2}$, where the finite-size anomalous points are extracted from $P({\cal D})$ and $\chi_M$. The horizontal dashed line marks the critical point from the scaling of the order parameter $M$ with the shaded area depicting its error. The error bar of the finite-size data (empty marker) depicts the uncertainty in $T$ determined by the minimum increment $dT=0.01$; the error bar for the extrapolated value (filled marker) is the standard deviation error from the curve fitting. The extrapolated $T^*_{c2}(\infty) = 0.930(5)$ and 0.938(6) for $D_0$ and $\chi_{|M|}$, respectively. (b) The susceptibility $\chi_{|M|}$ versus $T$ for different system sizes.}
\label{fig:T_star}
\end{figure}

Besides the scaling, one can also estimate the critical point from the finite-size extrapolation in case the finite-size critical points can be determined. As shown in Fig.~\ref{fig:Mu_A} (a), there is a shape jump before the peak center ${\cal D}_0$ reaches its uncorrelated value $0.5$ in the disordered phase, and an anomaly occurs for the peak height at the same temperature in Fig.~\ref{fig:Mu_A} (b). We consider the first point right after the jump as the finite-size critical point (marked as $T^*_{c2}$) and examine its extrapolated value in the thermodynamic limit. In the BTK transition, the finite-size shift of the critical point scales as $\propto 1/\ln^2(L)$~\cite{Sandvik2010}, thus in Fig.~\ref{fig:T_star}(a) we displayed the size-dependence of $T^*_{c2}$ in the same scale. This extrapolation successfully catches the critical point, and the extrapolated $T^*_{c2}(\infty)$ agrees surprisingly well with the $M$ scaling. We are then interested in whether one can take a similar extrapolation procedure from the order parameter. Here we extract the more distinct anomalous for the absolute value $|M|$ in Fig.~\ref{fig:ML}(a) from its susceptibility 
\begin{align}
\chi_{|M|}=\frac{\beta}{N}(\langle |M|^{2}\rangle-\langle |M|\rangle^{2}).
\label{eq:susc_abs}
\end{align}
The data of $\chi_{|M|}$ is displayed in Fig.~\ref{fig:T_star}(b), where the $T_{c2}^*(L)$ is determined by the position of the sharp peak. In Fig.~\ref{fig:T_star}(a), the extrapolation of the critical point from $\chi_{|M|}$ gives us the same $T_{c2}^*$ in the thermodynamic limit, as shown in Table~\ref{tab:Tc}. In summary, quantifying the MC procedure with the distribution of $\cal D$ catches the correct critical points, and the $P({\cal D})$ peak features a similar critical behavior with the order parameter and its susceptibility. 

\subsection{Intrinsic dimension}
\label{subsec:Id}

\begin{figure}[!t]
\includegraphics[width=\columnwidth]{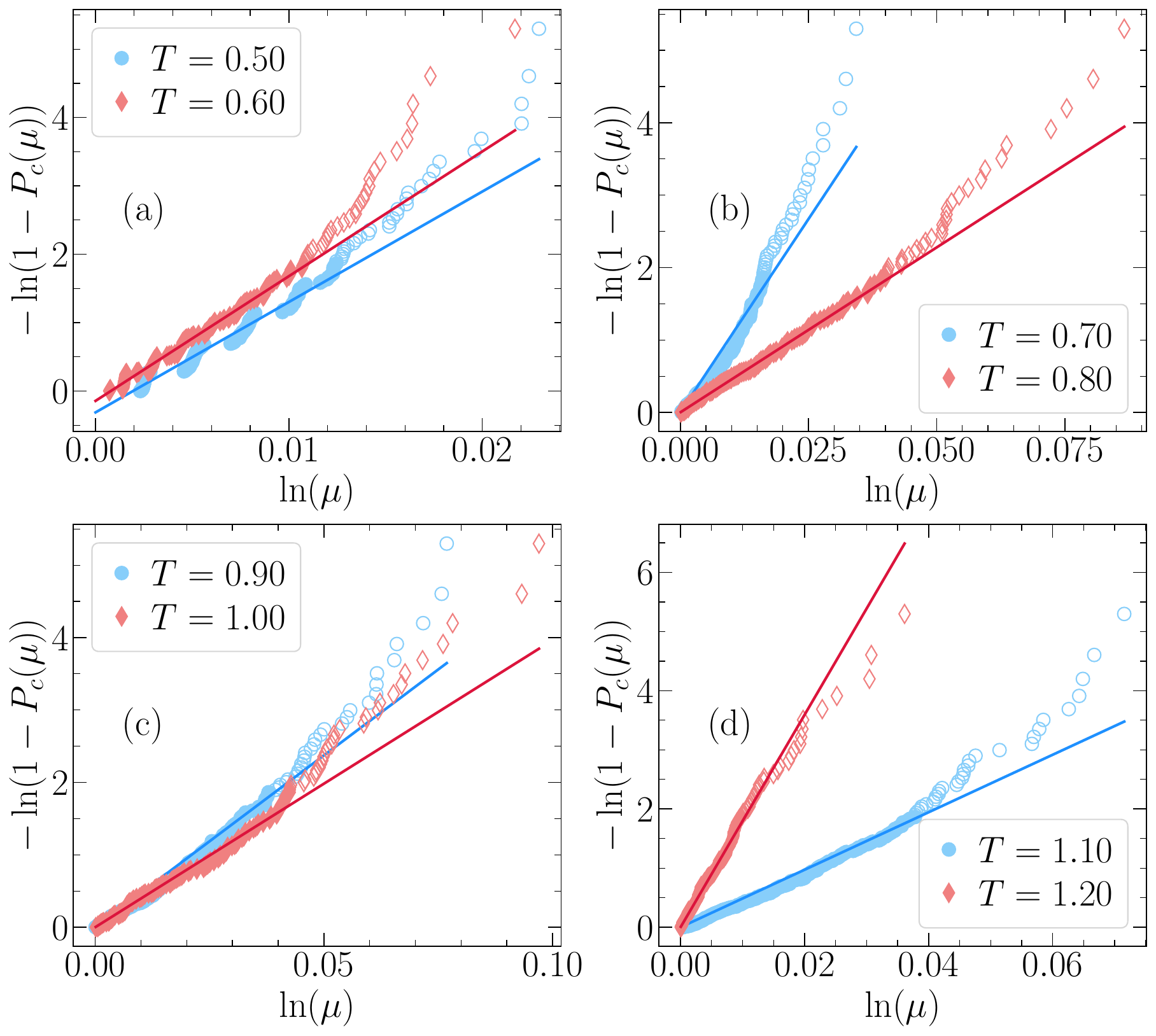}
\caption{The fitting process following Eq.~\eqref{eq:intrinsicd} to obtain the Intrinsic Dimension for a single MC bin. Panels (a-d) show the data for various increasing temperatures. The linear fitting adopts the first $85\%$ data for all temperatures, as depicted by the filled markers. }
\label{fig:get_Id}
\end{figure}

On the other hand, in some sense, the process of the MC sampling in the exponentially large configuration space shares a similar idea with the growing field of machine learning (ML), which identifies the universal property of the high-dimensional data sets from minimally processed data collection. Recently, ML ideas have motivated various applications in the context of statistical physics~\cite{melko,ising2,ising3,xy1,xy3,xy2,santos1,id2,xy4}. While most of the works focus on analyzing the dimension reduction results from other methods, recent works demonstrated that the reduction procedure could provide the same information~\cite{santos1,id2,xy4}. Among them, the authors of Ref.~\cite{santos1} and ~\cite{id2} employ the concept of the intrinsic dimension $I_d$, which roughly measures the minimum number of variables required to describe the global features of a large data set~\cite{goldt}, and compute $I_d$ of the MC thermal configurations to probe different kinds of phase transitions. 

In this section, we follow the same approach in Ref.~\cite{santos1} and compute the $I_d$ using the so-called two-NN method~\cite{Facco2017}, which focuses only on the distances to the NN and NNN of each element in the data set, assuming that these two points provide a uniform drawn from small enough $I_d$-dimensional hyperspheres. Specifically, the data set interested in the present work is a small fraction of MC thermal configurations. Similar to the $P({\cal D})$, we choose 200 uncorrelated configurations in each bin. And for each configuration, the two NNs determined by the smallest (non-zero) Hamming distances define a ratio $\mu={\cal D}_2/{\cal D}_1$ which obeys the following distribution $f(\mu) = I_d \mu^{-I_d-1}$. In terms of a cumulative distribution $P_c(\mu)$, the intrinsic dimension satisfyies
\begin{align}
I_{d}=-\frac{\mathrm{ln}[1-P_c(\mu)]}{\mathrm{ln}(\mu)},
\label{eq:intrinsicd}
\end{align}
then one can get $I_d$ from a linear fitting of this non-decreasing data collection. Different from $P({\cal D})$ in Sec.~\ref{subsec:PD} where configurations from all bins are considered as a single data set, here $I_d$ is computed in each bin as for the physical quantities. One reason is that different bins might produce possible degenerate configurations, which makes the nearest neighbors ill-defined. Even in a single bin, this degeneracy may appear at very low temperatures, and $\mathrm{ln}[1-P(\mu)]$ does not scale linearly as $\mathrm{ln}(\mu)$, as displayed in Fig.~\ref{fig:get_Id}(a). Nevertheless, we perform the same linear-fitting approach for all temperatures, as the linear behavior looks promising for temperatures of interest where the phase transitions occur, as shown in panels (b-d) for $T$ from 0.70 to 1.30 in the same figure. Repeating this fitting procedure for all MC bins, one can obtain the average $I_d$ and its statistical error. 

\begin{figure}[!t]
\includegraphics[width=\columnwidth]{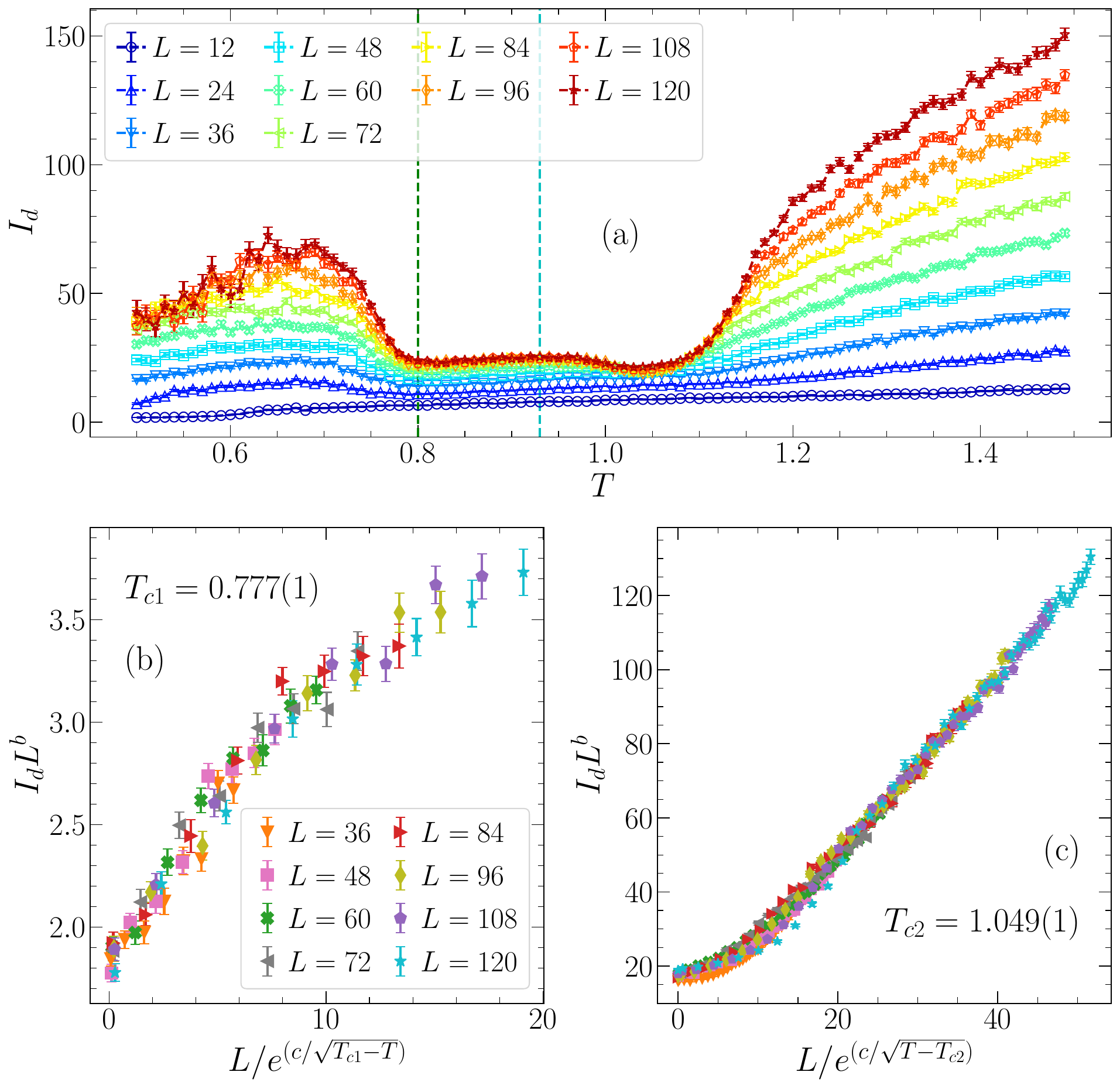}
\caption{(a) The intrisic dimenstion $I_d$ as a function of $T$ for different system sizes. (b) and (c) shows the data collapse around $T_{c1}$ and $T_{c2}$, respectively. We neglect data for $L=12$ and $24$ where $I_d$ is almost flat versus temperatures.The obtained non-universal constants are: (b) $b=-0.60(1)$, $c=0.51(1)$; (c) $b=-0.03$(1), $c=0.56(1)$. }
\label{fig:Id}
\end{figure}

The results of $I_d$ as a function of $T$ for different system sizes are displayed in Fig.~\ref{fig:Id}(a), where the curves at larger system sizes feature two anomalies that may correspond to the two critical points. For the first transition, the positions of the anomaly are close to $T_{c1}$, as for all other quantities examined in the previous text. However, the second anomaly is far from $T_{c2}$ and hardly varies for different $L$. Moreover, for smaller system sizes with $L=12$ and $24$, $I_d$ is almost flat with no apparent anomaly since $I_d$ may fail to catch the universal information for the relatively small data set~\cite{santos1}. Excluding data for the two smallest sizes, we perform the same scaling procedure as for the distribution of Hamming distances, with a BKT scaling form $I_d L^b={\cal F}(L/\xi)$ and three unknown parameters to be settled. However, as shown in Fig.~\ref{fig:Id}, the data collapse is not as good as previous quantities [See Fig.~\ref{fig:scaling_MXM} and~\ref{fig:A_scaling}], and the obtained critical points are inconsistent with the actual phase transitions, especially for the second critical point.

\subsection{Robustness and generality of $P({\cal D})$}
\label{sec:generity_PD}

\begin{figure}[!b]
\includegraphics[width=\columnwidth]{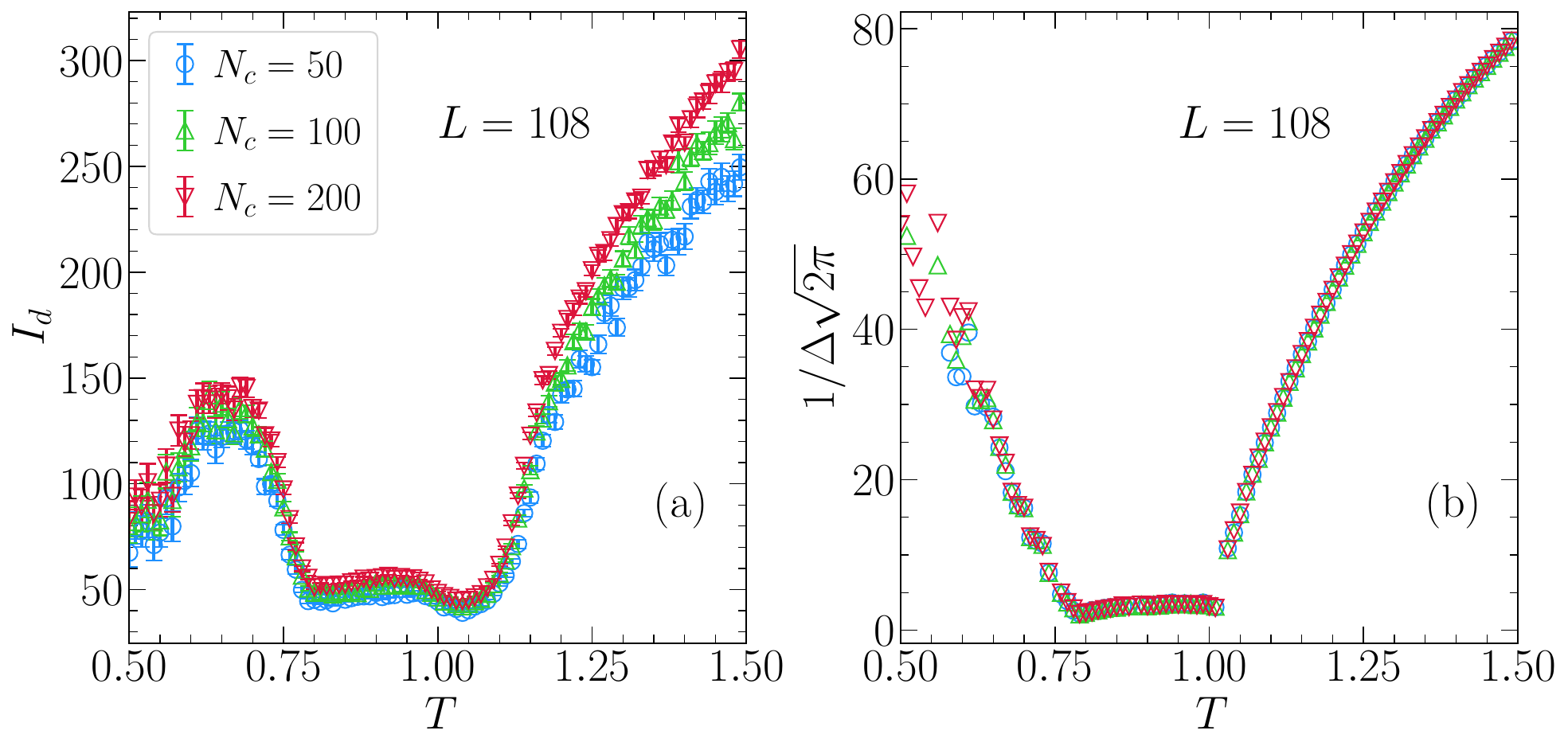}
\caption{The (a) intrinsic dimension $I_d$ and (b) effective height of $P({\cal D})$ versus temperature for different number of configurations $N_c$. }
\label{fig:test_Nc}
\end{figure}

While the origin of the failure in locating the critical points using the intrinsic dimension can be very complicated, some known disadvantages may affect the scaling. First, the method estimating $I_d$ is not unique, and the results can be different when using different approaches. Second, when using the two-NN method in the present work, $I_d$ is extracted from the linear fitting of Eq.~\eqref{eq:intrinsicd} with a selected portion of data that roughly satisfies the linear behavior. This procedure is not totally unbiased. Moreover, the value of $I_d$ sensitively depends on the number of configurations $N_c$ and does converge due to the so-called curse of dimensionality~\cite{santos1}. In contrast, the calculation of $P({\cal D})$ is very simple and robust. As displayed in Fig.~\ref{fig:test_Nc}, while $I_d$ varies as the number of configurations increases, $P({\cal D})$ is almost the same for different $N_c$. 

\begin{figure}[!b]
\includegraphics[width=0.8\columnwidth]{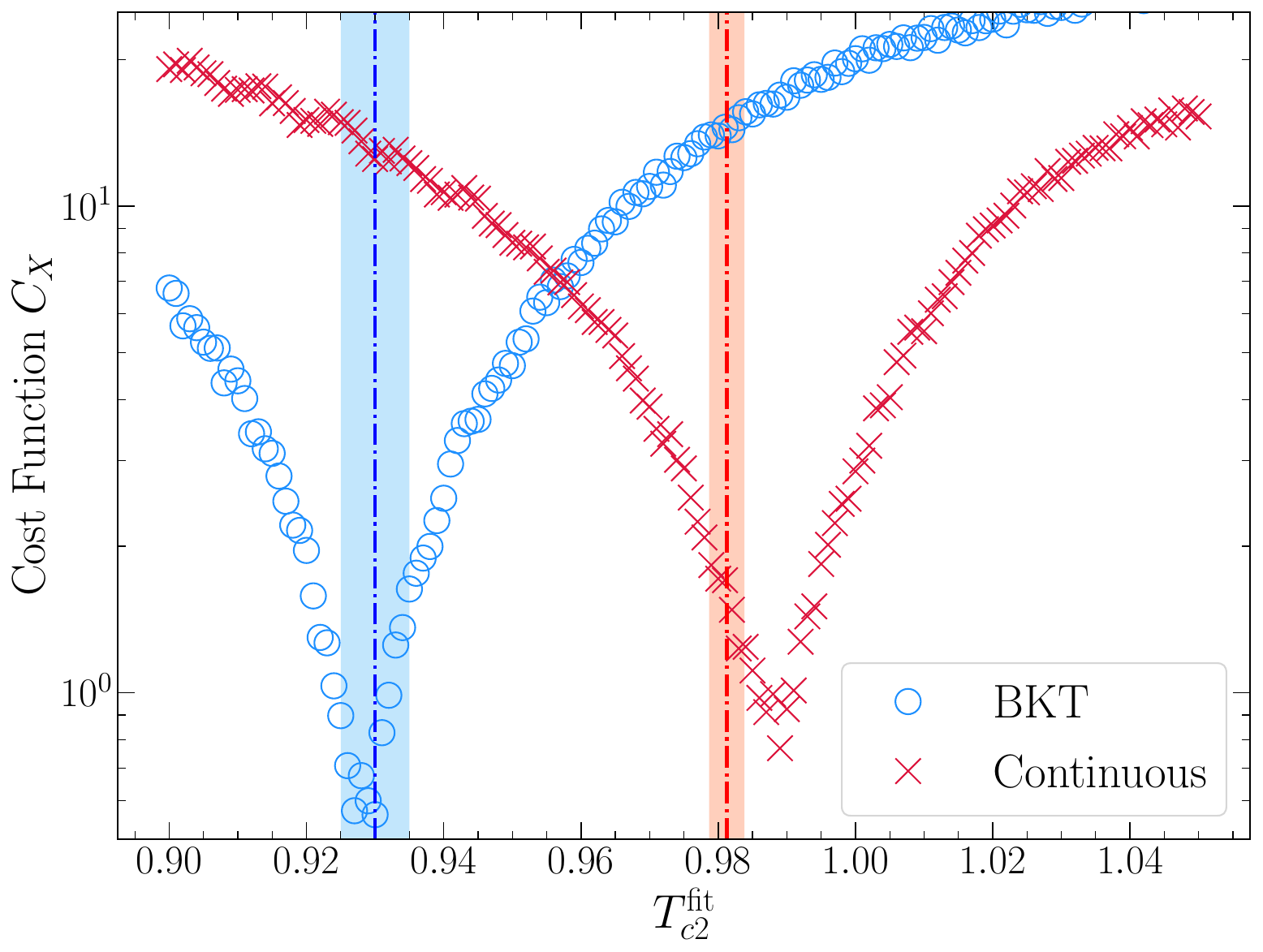}
\caption{Comparison of the cost functions $C_X$ [See Eq.~\eqref{eq:cost_function} and Appendix~\ref{appendix:data_collapse}] between scaling procedures with BKT and continuous form for $T_{c2}$. The vertical dashed line depicts the critical point from the finite size extrapolation, with the shading area depicting the error. Blue and red denote the BKT and continuous phase transitions, respectively. }
\label{fig:compare_con_BKT}
\end{figure}

\begin{table*}[!t]
    \centering
    \begin{tabular}{c|c|c|c|c|c|c}
        \hline
        \multicolumn{1}{c}{} &
        \multicolumn{4}{|c|}{BKT form scaling} &
        \multicolumn{2}{c}{finite-size extrapolation} \\
        \hline
                & $M$ & $\chi_M$  & $P({\cal D})$ peak height & $I_d$ & $P({\cal D})$ peak position/height & $\chi_{|M|}$  \\
        \hline
        $T_{c1}$ & ~0.803(5)~ & ~0.808(1)~ & ~0.806(1)~ & ~0.777(1)~ &  -        & -          \\
        \hline
        $T_{c2}$ & ~0.926(7)~ & ~0.929(1)~ & ~0.930(1)~ & ~1.049(1)~ &  ~0.930(5)~ & ~0.938(6)~   \\
        \hline
    \end{tabular}
    \caption{The critical points obtained from different quantities and approaches. The last two columns present $T_c^*$ from the finite-size extrapolation (See Fig.~\ref{fig:T_star}). Except for $I_d$, all other results show good consistency. }
    \label{tab:Tc}
\end{table*}

It is also interesting to consider the possibility of determining the critical point and transition type of the phase transitions by $P({\cal D})$ without knowing the order parameter. As shown in Fig.~\ref{fig:lattice}(b) and (c), the temperature dependence of the specific heat and the distribution of Hamming distances suggests three phases and two phase transitions. Besides, from the temperature dependence of the specific heat in Fig.~\ref{fig:lattice}(b) for different system sizes, one can rule out the first-order phase transition. As for the Hamming distance, at sufficiently high temperatures, the narrow Gaussian peak centering at ${\cal D}=0.5$ indicates uncorrelated configurations and a disordered phase. In the presence of any order of symmetry, the corresponding constraints in the configuration space build up correlations between configurations and drive $P({\cal D})$ away from the disordered limit. According to the Landau-Ginzburg-Wilson paradigm, the phase transition between two ordered phases should not be continuous unless all the variables are well-tuned~\cite{sachdev04,masuggest}, so it is reasonable to predict a BKT phase transition for the first critical point. The second one, from the ordered to the disordered phase, can be either a continuous or BKT phase transition. In this case, one can try the scaling procedure for both transition types and the numerics would favor the correct scaling form. 

Similar to the BKT scaling in Eq.~\eqref{eq:scaling_A}, we next perform the scaling that follows the continuous phase transition with the scaling form
\begin{align}
    \frac{1}{\Delta\sqrt{2\pi}}L^{b^*}={\cal F}_\mathrm{con}(t^*L^{1/\nu}),
    \label{eq:scaling_second}
\end{align}
where $b^*$ and $\nu$ are unknown critical exponent with $t^*=(T-T_c)/T_c$. The critical parameters can be determined by minimizing the cost function $C_X$ [See Eq.~\eqref{eq:cost_function} and Appendix~\ref{appendix:data_collapse}] in the three-dimensional parameter space $\{T_{c2},b^*,\nu\}$. To directly compare the two scaling forms, we fix the other two scaling parameters [$b^*$ and $\nu$ in Eq.~\eqref{eq:scaling_second}; $b$ and $c$ in Eq.~\eqref{eq:scaling_A}] for the best data collapse and display $C_X$ as a function of the test $T_{c2}$. As displayed in Fig.~\ref{fig:compare_con_BKT}, the BKT scaling form at the best data collapse has a smaller cost function. Another way of estimating $T_{c2}$ is to extrapolate the finite-size critical points to the thermodynamic limit, which scales as $\propto L^{1/\nu}$ for the continuous phase transition~\cite{Sandvik2010}. Using the $\nu$ obtained from scaling, we carry out a similar extrapolation as the process in Fig.~\ref{fig:T_star} and plot the obtained $T_{c2}^*$ in Fig.~\ref{fig:compare_con_BKT}. Compared to the BKT phase transition, the two critical points obtained by the scaling and extrapolation do not match well. Thus, the numerical experiments suggest a BKT phase transition, and the critical point can be achieved by finite-size scaling and extrapolation.

\section{Summary and discussion}
\label{sec:sum}

We numerically study the KIM and focus on its two consecutive BKT phase transitions with comprehensive classical MC simulations. After reinvestigating the phases and phase transitions employing the physical quantities such as the magnetic order parameter and its susceptibility, we propose that the BKT phase transitions can also be characterized by the information extracted from MC procedures in a new way that can be used in many different models and phase transitions. Specifically, we first select a small set of uncorrelated configurations determined by the MC procedure, then measure all Hamming distances between every pair of collections and make a target data collection. We demonstrate that the distribution of Hamming distances $P({\cal D})$ contains all information on the phase transitions, and the critical points can be extracted using the same BKT scaling form for the effective height of $P({\cal D})$ peak. Moreover, using anomaly in either height or position of the $P({\cal D})$ peak, one can successfully obtain the transition point by a finite-size extrapolation. 

We also compute the intrinsic dimension $I_d$ from the nearest neighbors of the selected MC configurations, which was recently used to analyze the phase transitions in the context of machine learning. However, compared to $P({\cal D})$, the value of $I_d$ is unstable at low temperatures, and the data collapse procedure does not work well for both phase transitions. For comparison and summary, we list the two critical points obtained from different approaches in Table.~\ref{tab:Tc}. As one can see, while from $P({\cal D})$ one obtains the critical points with surprisingly high accuracy, $I_d$ fails to catch the correct phase transitions.

\begin{figure*}[!t]
\includegraphics[width=1\linewidth]{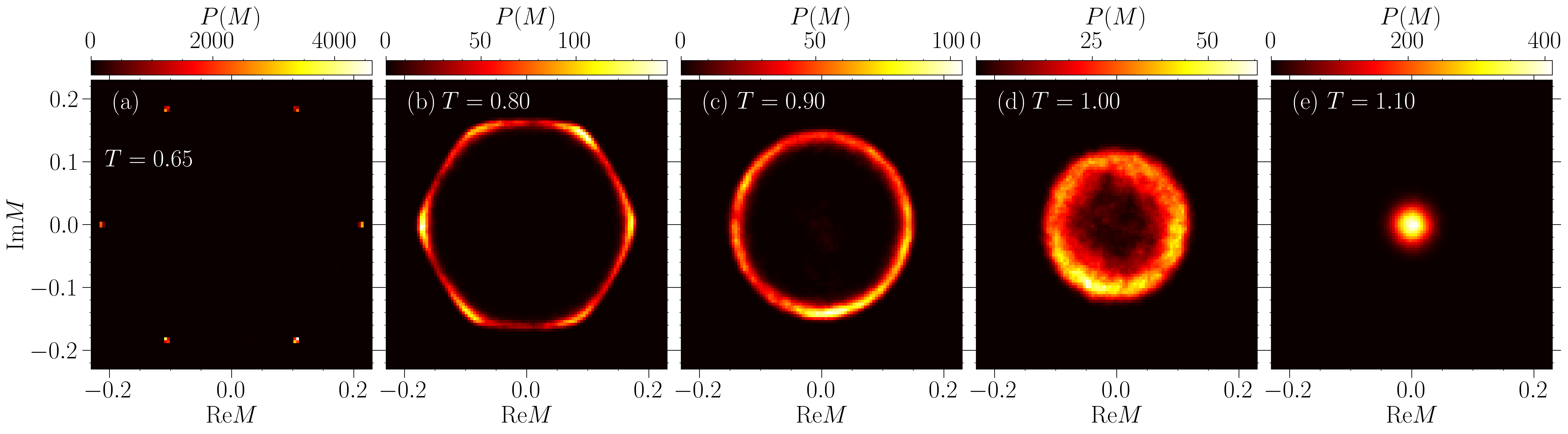}
\caption{The histogram of the order parameter $M$ in the complex \{Re$M$, Im$M$\} plane for $L=120$ and different temperature: (a) $T=0.65$, (b) $T=0.80$, (c) $T=0.90$, (d) $T=1.00$ and (e) $T=1.10$. The histogram is carried out for all MC configurations in 56 independent MC bins.}
\label{fig:M_hist} 
\end{figure*}

Quantifying the MC process rather than the results to identify the phase transition has attracted increasing interest in recent years, especially in cooperation with ML ideas. Inspired by these previous investigations, we have proposed a rather simple but robust approach to tackle BKT phase transitions in KIM. Our method not only accurately catches the critical points based on the known scaling form but can also settle the type of transition by the numerical experiment without knowing the order parameter, which allows tackling the phase transition for systems with less knowledge. One immediate example is the compound HoAgGe claimed to support the Kagome spin-ice ground state in experiment~\cite{KZhao}. The system is described by an extended KIM with third-nearest-neighbor couplings, long-range dipolar interactions, and lattice distortion. According to the temperature dependency of specific heat from the experimental measurements and a rough computing analysis, the authors of Ref.~\cite{KZhao} suggest two intermedia phases between the low-temperature spin-ice and high-temperature disordered phase as the temperature increases. However, the knowledge about the intermediate phases and all three phase transitions is still poor. Our proposal can be naturally employed in the extended KIM to get a better understanding of the system.

On the other hand, our method can be useful for the quantum spin liquid without a well-defined order parameter~\cite{spinliquid}, or quantum systems with severe sign problems where quantum MC simulation cannot give us meaningful physical quantities because of the negative weight on the sampled configurations. In these cases, extracting the information on phase transitions directly from the configurations is especially important. Although based on the classical MC simulation, our findings can be useful in studying quantum systems with the same protocol as long as the MC simulation is carried out on a configuration basis. Of course, the quantum system with frustrations is much more intractable, where the MC sampling itself can be unstable. We hope the present results can shed some light on these challenges, and the research along this line deserves further investigation.

\section*{Acknowledgments}
N. M. thanks Kan Zhao for discussions and collaborations in related contributions. This research was supported by the National Natural Science Foundation of China (grant nos. 12004020, 12174167, 12247101), the 111 Project under Grant No.20063, and the Fundamental Research Funds for the Central Universities.

\appendix

\section{Order parameter in the complex plane}
\label{sec:M_hist}

The main text focuses mainly on the two BKT phase transitions and the corresponding scaling approach to the critical points. Here we present our results on the phases characterized by the distribution of $M=|M|e^{{\mathrm i}\phi}$ in the complex plane \{Re$M$, Im$M$\} at different temperatures. At low temperatures, the system features the same phase as the ground state, where $|M|=2/9$ is constant, and $\phi$ only takes six fixed values. As displayed in Fig.~\ref{fig:M_hist}(a) for numerical results at a low temperature $T=0.65$, which agrees with the schematic pattern for the ground state in Fig.~\ref{fig:six-clock}. When temperature increases, a $U(1)$ symmetry emerges with a nonzero but temperature-dependent $|M|$, as exhibited by the circle in the complex plane \{Re$M$, Im$M$\} in panel (c) for $T=0.90$. All orders break down for sufficiently large $T$ due to strong thermal fluctuations with the order parameter $|M|=0$, and the system is in the disordered phase. For finite system sizes, as shown in Fig.~\ref{fig:M_hist}(b) and (d), one also observes some intermediate patterns for the $M$ distribution, which will disappear in the thermodynamic limit. For all temperatures, $M$ distribution is centrally symmetric of zero. 

\section{Hamming distance of the six-clock ground state and multi-Gaussian-peak fitting}
\label{appendix:six-clock}

\begin{figure}[!b]
\includegraphics[width=\columnwidth]{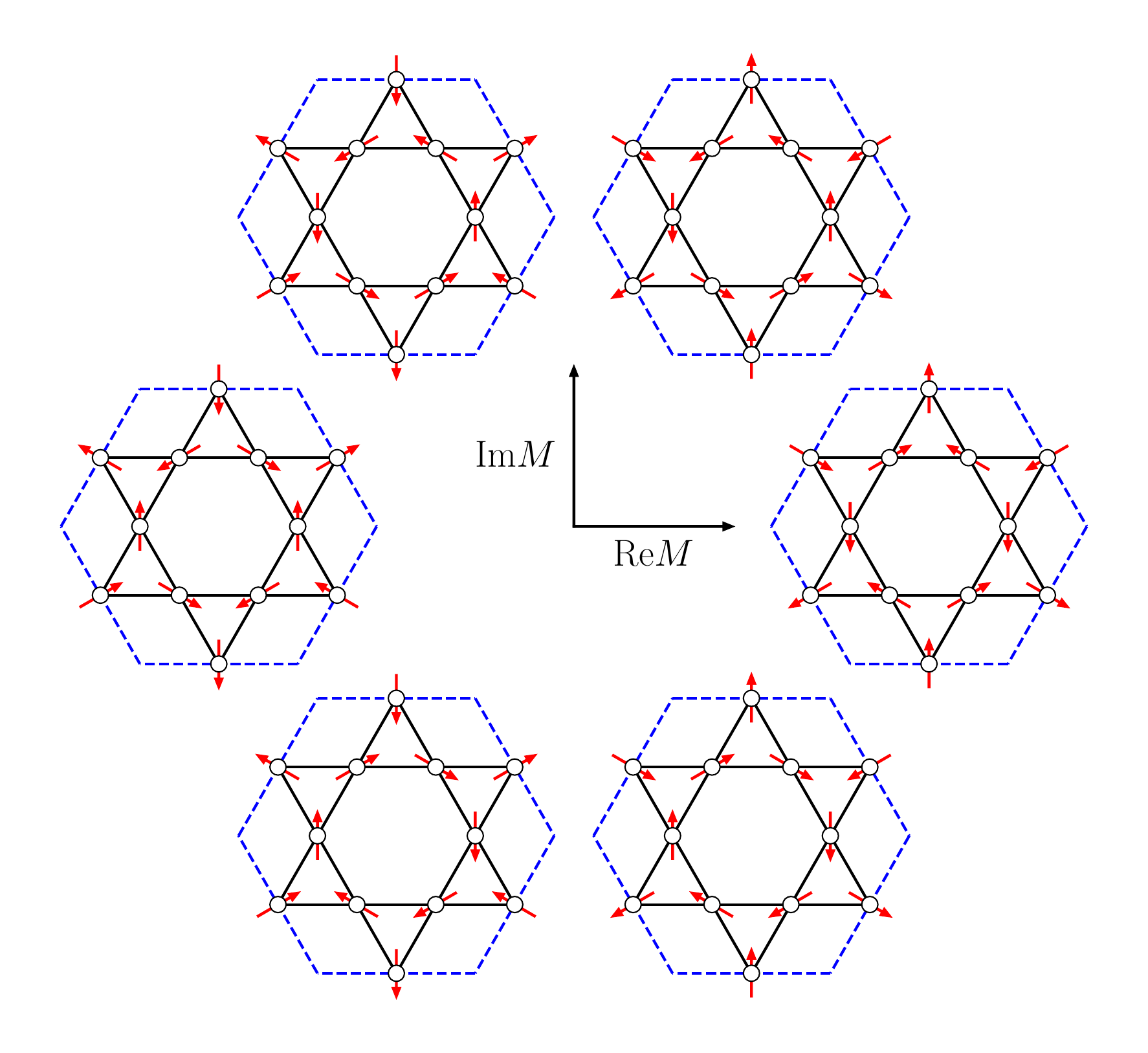}
\caption{The schematic plot of the six-clock spin ice ground state by spin configurations in the super unit cell of 9 spins (6 spins on boundaries are shared with other super unit cells). Here the arrow at site $i$ pointing towards (away from) the center of the unit cell indicate $\sigma_i=1$ (-1). The Hamming distance ${\cal D} = 1/3$, $2/3$ and $1$ between super unit cells at NN, NNN, and opposite positions.}
\label{fig:six-clock}
\end{figure}

The KIM has the ground state of six-clock state spin ice, as explicitly demonstrated in Fig~\ref{fig:six-clock}. The six-fold degenerate ground states satisfy $M=|M|e^{\mathrm{i}\phi}$, with $\phi=n\pi/3$ and $n=0,1,2,3,4,5$. All possible Hamming distances can be easily obtained by counting the different spins between these six degenerate spin configurations. For example, considering the state with $n=0$ and $1$, the three different spins between them result in a Hamming distance ${\cal D}=1/3$ as there are in total nine spins in the super unit cell. Repeating this counting procedure, it is easy to find out that all NN pairs of super unit cells have ${\cal D}=1/3$, all NNN pairs have ${\cal D}=2/3$, and the opposite pairs have ${\cal D}=1$. Taking into account that the MC steps around one of the configurations at low temperatures, which is metastable in MC procedures and results in configurations with ${\cal D}=0$, all possible $\cal D$ explain the sharp peaks and their positions at low temperatures in Fig.~\ref{fig:lattice}(c) and \ref{fig:PD}.

Not only the possible values of $\cal D$, we can also estimate the weight of $P({\cal D})$ at low temperatures. In the data set of uncorrelated MC thermal configurations, all six degenerate ground states have the same weight, with a total of 36 possible combinations. The number of combinations supporting ${\cal D}=0$ ($1$) is $6$, resulting a $P({\cal D})=1/6$; Each states have two NNs and NNNs, thus $P({\cal D})=1/3$ for ${\cal D}=1/3$ and $2/3$. Despite the thermal fluctuations that broaden the delta peak, the peak heights of $P({\cal D})$ in Fig.~\ref{fig:PD} at low temperatures confirm our estimation. 

For the six-clock spin ice ground state in the present work, we already know the number of peaks. To obtain the peak information (the position and height/width), it is natural to fit the data using a two-Gaussian form, where the Gaussian broadening considers the finite temperature effect. In general, as discussed before in this subsection, for the $q$-state clock model with any $q$, one can predict the number of peaks in $P(D)$ and carry out a similar procedure. However, for generic unknown systems, it may be better to perform an unbiased numerical analysis to extract distinct information of $P(D)$, such as the number of data centers, the center average values, and the variance of each data center.

\section{Data collapse and uncertainty}
\label{appendix:data_collapse}

\begin{figure}[!b]
\includegraphics[width=\columnwidth]{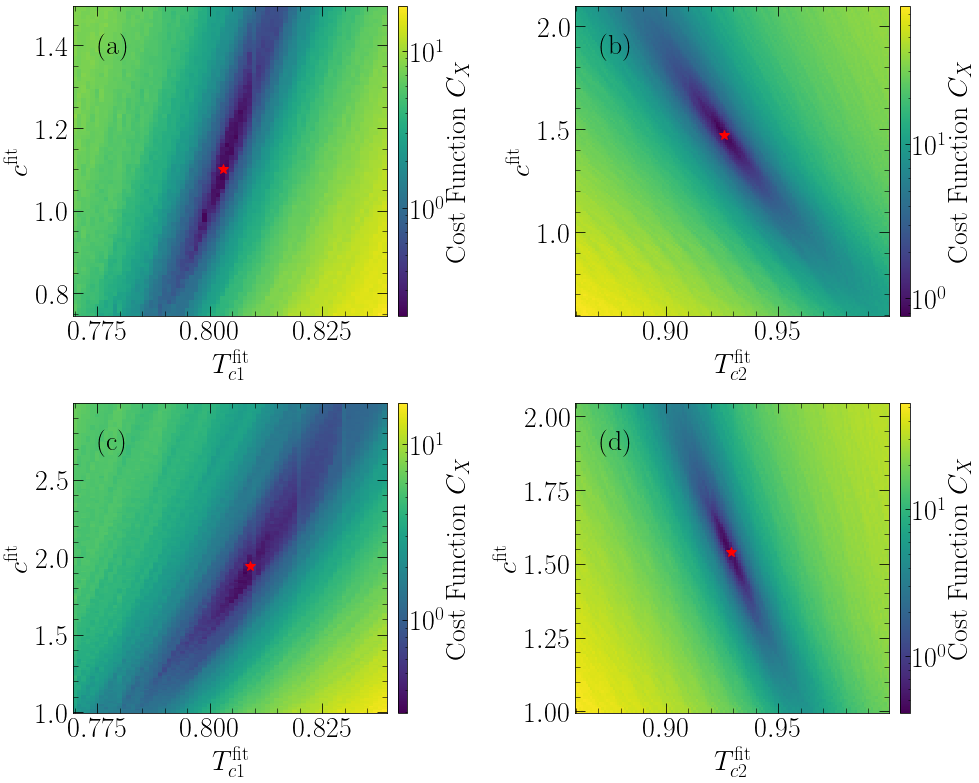}
\caption{The cost function $C_X$ in the two-dimensional parameter space $\{T_{c},c\}$, for the scaling of $T_{c1}$ ($T_{c2}$) in left (right) panels and $M$ ($\chi_M$) in upper (lower) panels. The minimum is marked by the red star, with (a) $T_{c1}=0.803$, $c=1.16$, (b) $T_{c2}=0.926$, $c=1.47$, (c) $T_{c1}=0.808$, $c=1.94$, and (d) $T_{c2}=0.929$, $c=1.54$. 
}
\label{fig:scaling_error}
\end{figure}

In this work we obtain the critical $T_c$, $\eta$, and $c$ in the scaling form of Eq.~\eqref{eq:scaling2} from the best data collapse with the minimum error defined by the following cost function~\cite{Jan2020,Aramthottil2021,Mondaini2022,liang2023disorder}
\begin{align}
    C_X=\frac{\sum_j |X_{j+1}-X_j|}{\max\{X_j\}-\min\{X_j\}}-1,
    \label{eq:cost_function}
\end{align}
where $C_X$ is a data collection of all $M(L,T)$ [$\chi_M(L,T)$] values in the parameter space for different temperature and system sizes. After sorting $X_j$ in a nondecreasing way with $X_j\leq X_{j+1}$, the minimum $C_X$ gives the smoothest curve of all collected data. Since $\eta$ at the critical point is known, we fix $\eta=1/9$ ($1/4$) in solving the first (second) critical points with two in the two-dimensional parameter space $\{T_{c},c \}$. 

In practice, one obtains a parameter-dependent cost function $C_X(T_{c},c)$ for each pair of fitted parameters value, and repeating this procedure in the two-dimensional parameter space $\{T_{c},c\}$ gives us the minimum of $C_X$ and the best data collapse. As shown in Fig.~\ref{fig:scaling_error}, the cost function displays a unimodal function in the target region of the $\{T_{c},c\}$ plane for both two critical points and physical quantities. Therefore, we can easily extract the unambiguous minimum of $C_X$ and get the corresponding data collapse in Fig.~\ref{fig:scaling_MXM} in the main text. Note that all three parameters are unknown for the scaling of the $P(\cal D)$ height and the intrinsic dimension $I_d$, and the minimization in Fig.~\ref{fig:A_scaling} and \ref{fig:Id} are carried out in three dimensions. 

Equation~\eqref{eq:cost_function} and the above procedure do not provide the uncertainty or error of the target parameters. Here the uncertainty is estimated by performing three data collapses, with the data collection of $X_j$, $X_j+\sigma_{X_j}$ and $X_j-\sigma_{X_j}$, and the corresponding obtained critical temperature denoted as $T_c$, $T_c^+$ and $T_c^-$. Then the error of $T_c$ is defined as $\sigma_{T_c}=\mathrm{max}(|T_c^+-T_c|,|T_c^- - T_c|)$, and in the main text [See Fig.~\ref{fig:scaling_MXM} as an example] the critical temperature is expressed as $T_c\pm \sigma_{T_c}$. Other target parameters in scaling follow the same procedure and expression. 

\section{Simulation Benchmark with Ref.~\cite{chern}}
\label{appendix:benchmark}

\begin{figure}[!t]
\includegraphics[width=\columnwidth]{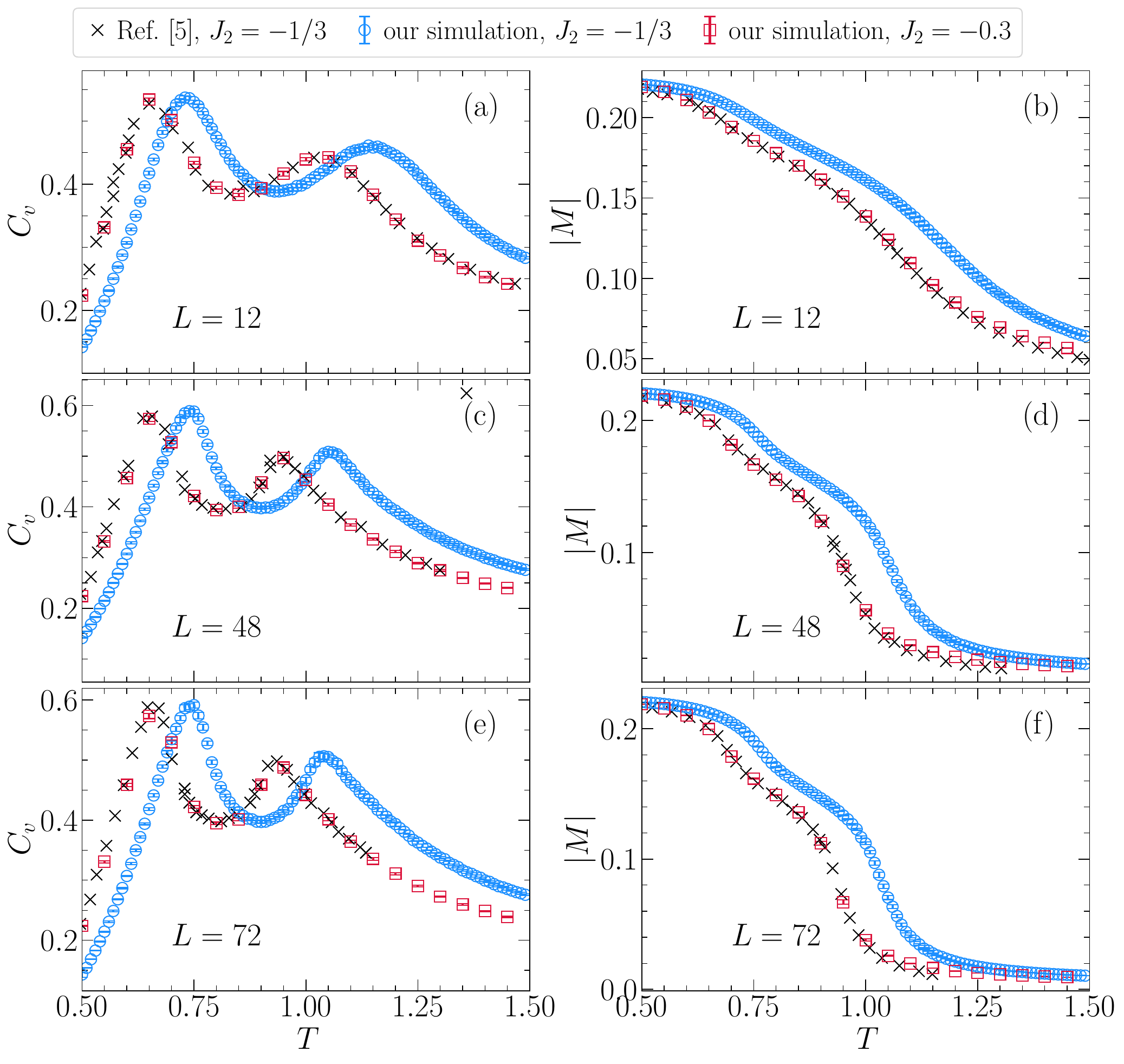}
\caption{Comparison of physical quantities (specific heat $C_v$ and order parameter $|M|$) between the present work and Ref.~\cite{chern}. The results of Ref.~\cite{chern} while claiming $J=-1/3$ agree extremely well with our simulations of $J=-0.3$. }
\label{fig:benchmark}
\end{figure}

One would notice that the critical points in the present work are larger than what is obtained in Ref.~\cite{chern}, even though the two works use the same model, parameters, and numerical approaches. However, except for the quantitative difference, the results in both works are quite self-consistent and support the same physics. It is unlikely that either work has made essential mistakes in the model and numerical simulations. As displayed in Fig.~\ref{fig:benchmark}, the direct comparison between our simulations with the results in Ref.~\cite{chern} shows apparent disagreement at finite system sizes, so the simple finite size effect cannot explain the discrepancy as well. 

After carefully checking our calculation and assuming both works have obtained correct and physically sound results, an intuitive guess is that we may use the different parameter $J_2$. Considering that the value $1/3$ is sometimes replaced by the approximation with finite precision in practical numerics, we perform some test calculations of several approximated values (such as 0.3 and 0.33) and eventually found a perfect match with results in Ref.~\cite{chern} with $J_2=-0.3$, as displayed in Fig.~\ref{fig:benchmark}. Either on purpose or by accident, Ref.~\cite{chern} took an approximate value of $0.3$ while claiming $1/3$ in their paper, which however does not affect the novelty of their comprehensive investigations.

\bibliography{main}

\end{document}